\documentclass[structabstract]{aa}
\usepackage{natbib}        
\bibpunct{(}{)}{;}{a}{}{,} 
\usepackage{tabularx}
\usepackage{amsmath}       
\usepackage{multirow}
\usepackage[normalem]{ulem}

\usepackage{graphicx}
\usepackage{txfonts}

\newcommand {\mmean}[1] {\left<\textstyle{#1}\right>}
\newcommand {\nbody}    {\textsc{\mbox{Nbody6}}}

\newcommand {\Msun}     {\mbox{M$_{\odot}$}}

\newcommand {\lmst}     {\mbox{$l_{\mathrm{MST}}$}}
\newcommand {\lref}     {\mbox{$l_{\mathrm{MST}}^{\mathrm{ref}}$}}
\newcommand {\lmass}    {\mbox{$l_{\mathrm{MST}}^{\mathrm{mass}}$}}
\newcommand {\Lmst}     {\mbox{$\Lambda_{\mathrm{MST}}$}}
\newcommand {\gmst}     {\mbox{$\gamma_{\mathrm{MST}}$}}
\newcommand {\gref}     {\mbox{$\gamma_{\mathrm{MST}}^{\mathrm{ref}}$}}
\newcommand {\gmass}    {\mbox{$\gamma_{\mathrm{MST}}^{\mathrm{mass}}$}}
\newcommand {\Gmst}     {\mbox{$\Gamma_{\mathrm{MST}}$}}
\newcommand {\Mmfd}     {\mbox{${\cal{M}}_{\mathrm{MF}}^{\mathrm{d}}$}}
\newcommand {\Mmfc}     {\mbox{${\cal{M}}_{\mathrm{MF}}^{\mathrm{c}}$}}
\newcommand {\Mrch}     {\mbox{${\cal{M}}_{\mathrm{R}}^{\mathrm{ch}}$}}

\newcommand {\Mmstl}    {\mbox{${\cal{M}}_{\mathrm{MST}}^{\mathrm{\Lambda}}$}}
\newcommand {\Mmstg}    {\mbox{${\cal{M}}_{\mathrm{MST}}^{\mathrm{\Gamma}}$}}

\begin{document}
\title{A highly efficient measure of mass segregation in star clusters}


   \author{C. Olczak\inst{1,2,3,4}
     \and R. Spurzem\inst{3,1,4}
     \and Th. Henning\inst{2}}

   \institute{
     Astronomisches Rechen-Institut (ARI), Zentrum f{\"u}r Astronomie Universit{\"a}t Heidelberg, M{\"o}nchhofstrasse 12-14, 69120 Heidelberg,
     Germany \\  \email{olczak@ari.uni-heidelberg.de}
     \and 
     Max-Planck-Institut f{\"u}r Astronomie (MPIA), K{\"o}nigstuhl 17, 69117 Heidelberg, Germany
     \and
     National Astronomical Observatories of China, Chinese Academy of Sciences (NAOC/CAS), 20A Datun Lu, Chaoyang District, Beijing 100012,
     China
     \and The Kavli Institute for Astronomy and Astrophysics at Peking University (KIAA), Yi He Yuan Lu 5, Hai Dian Qu, Beijing 100871, China
   }

   \date{Received ; accepted}

%
%

  \abstract
  {Investigations of mass segregation are of vital interest for the understanding of the formation and dynamical evolution of stellar systems on a
    wide range of spatial scales. A consistent analysis requires a robust measure among different objects and well-defined comparison with theoretical
    expectations. Various methods have been used for this purpose but usually with limited significance, quantifiability, and application to both
    simulations and observations.}
  {We aim at developing a measure of mass segregation with as few parameters as possible, robustness against peculiar configurations, independence of
    mass determination, simple implementation, stable algorithm, and that is equally well adoptable for data from either simulations or observations.}
  {Our method is based on the minimum spanning tree (MST) that serves as a geometry-independent measure of concentration. Compared to previous such
    approaches we obtain a significant refinement by using the geometrical mean as an intermediate-pass.}
  {The geometrical mean boosts the sensitivity compared to previous applications of the MST. It thus allows the detection of mass segregation with
    much higher confidence and for much lower degrees of mass segregation than other approaches. The method shows in particular very clear signatures
    even when applied to small subsets of the entire population. We confirm with high significance strong mass segregation of the five most massive
    stars in the Orion Nebula Cluster (ONC).}
  {Our method is the most sensitive general measure of mass segregation so far and provides robust results for both data from simulations and
    observations. As such it is ideally suited for tracking mass segregation in young star clusters and to investigate the long standing paradigm of
    primordial mass segregation by comparison of simulations and observations.}


   \maketitle

%

\section{Introduction}

\label{sec:introduction}

It is commonly accepted that star formation does usually not occur in isolation but that a large majority of young stars -- up to 90\,\% -- are part
of a cluster \citep{2003ARA&A..41...57L,2009ApJS..181..321E}. The dynamical evolution of a star cluster leaves a variety of imprints in the phase
space of its stellar population which are good tracers of the \emph{dynamical} age of the cluster. This quantity is in particular interesting when
compared to the \emph{physical} age. A higher dynamical than physical age means that observable dynamical imprints did not have enough time to evolve
dynamically and thus must have been present - at least partially - already at the beginning. This is usually known as \emph{primordial} origin.

One of the most widely discussed aspects of the dynamical evolution of young star clusters is that of mass segregation. From theoretical work it is
well known that this process is inevitably entangled with the dynamical evolution of a self-gravitating system of at least two different mass
components \citep{1969ApJ...158L.139S,1983ApJ...271...11F,1995MNRAS.272..772S,2007MNRAS.374..703K}. Due to energy equipartition -- hence via two-body
encounters -- the more massive particles tend to settle towards the cluster centre over time while the lower-mass particles are preferentially pushed
to the outer parts. However, it is a much more challenging task to identify mass segregation observationally in real objects than theoretically from
`clean' numerical simulations.  This is even more severe for young star clusters that are usually still embedded in their natal gas and the dynamical
and physical age of which is much more difficult to estimate.

However, the investigation of mass segregation in these young stellar systems is of particular interest for a deeper understanding of the star
formation process. Three fundamental questions are part of the scientific discussion in this context: (i) Do young star clusters (really) show mass
segregation? (ii) What is the observed degree of mass segregation? (iii) Could the observed degree of mass segregation have developed dynamically or
can it be only explained by primordial origin?

An investigation of these important aspects of the star formation process requires a solid tool that is at best independent of the method used to
determine the stellar masses and independent of the geometry of the object, that provides an unambiguous measure and is equally well applicable to
observational and numerical data. Note that also dynamical models have the equivalent problem to identify mass segregation clearly and quantitatively.

So far mainly four measures have been used to investigate mass segregation in (young) star clusters: i) the slope of the (differential) mass function
in different annuli around the cluster centre \citep[$\Mmfd$; e.g.][]{1988ApJ...329..187R,1989ApJ...341..168B,1997AJ....113.1733H}, ii) the slope of
the cumulative mass function in different annuli around the cluster centre \citep[$\Mmfc$; e.g.][]{1992BASI...20..287P,1998ApJ...492..540H}, iii) the
characteristic radius of different mass-groups of stars \citep[$\Mrch$; e.g.][]{1983ApJ...271...11F}, and iv) the length of the minimum spanning tree
(MST) of different mass-groups \citep[$\Mmstl$; developed by][]{2009MNRAS.395.1449A}\footnote{A variant of the MST method has been presented just
  recently by \citet{2011arXiv1103.0406Y} at the time of submission of our publication.}.

Most of these methods suffer from several weaknesses. The first three, $\Mmfd$, $\Mmfc$, and $\Mrch$, implicitly assume a spherical geometry and thus
depend on the determination of some cluster centre. The first two of them introduce additional bias due to radial binning and uncertainties in
deriving the slope of the mass function. Furthermore, $\Mmfd$ suffers from uncertainties due to mass binning \citep[see e.g.][for a comparison of
$\Mmfd$ and $\Mmfc$ applied to observational data]{2006AJ....132..253S}. There is a fundamental difference in the concept of the first and the last
two methods: the former measure the mass distribution in different spatial volumes, the latter evaluate the spatial distribution of different sets of
most massive stars. Consequently, $\Mrch$ and $\Mmstl$ do \emph{not} require a direct measure of stellar masses but only a qualitative criterion for
correct ordering. This property is a huge advantage in the face of observational data. However, $\Mrch$ has also a significant weakness: the
characteristic radius of a small subgroup is very sensitive to the definition of the cluster centre. Hence this method is not viable for a low degree
of mass segregation, i.e. a signature from only a few most massive stars.

In contrast, $\Mmstl$ does not show any of these disadvantages. However, it is also affected by quite low sensitivity that prevents unambiguous
detection of weak mass segregation. To take advantage of the potential strength of $\Mmstl$ we have developed an improved prescription for measuring
mass segregation, in the following referred to as $\Mmstg$, with significantly increased sensitivity. We present in this work our scheme and
demonstrate its efficiency.

In Section~\ref{sec:methods} we describe our method $\Mmstg$ for measuring mass segregation and discuss a scheme for setting up initially mass
segregated star cluster models. In Section~\ref{sec:results} we discuss various examples of the effectiveness of our scheme compared to previous
methods and present a first numerical application. The conclusion and discussion mark the last section of this paper.

%

\section{Methods}

\label{sec:methods}

\subsection{Geometrical minimum spanning tree $\Gmst$}

\label{sec:methods:mst}

As a proxy for mass segregation we extend the method $\Mmstl$ developed by \citet{2009MNRAS.395.1449A} \citep[see
also][]{2004MNRAS.348..589C,2006A&A...449..151S}. In summary, the authors use the minimum spanning tree (MST), the graph which connects all vertices
within a given sample with the lowest possible sum of edges and no closed loops \citep{1969JRSS...18...54G}. The length of the MST, $\lmst$, is a
measure of the concentration or compactness of a given sample of vertices and is unique whereas its shape does not need to be. Mass segregation of a
stellar system of size $N$ is quantified by comparing $\lmst$ of the $n$ most massive stars, $\lmass$, with the average $\lmst$ of $k$ sets of $n$
random cluster stars, $<\lref>$, and its standard deviation, $\Delta\lref$. The distribution of $\lmst$ of the $k$ random samples is indeed
Gaussian. The ratios of these quantities,
\begin{equation}
  \Lmst       = \frac{ <\lref> }{ \lmass } \,, \quad
  \Delta\Lmst = \frac{ \Delta\lref }{ \lmass } \,,
  \label{eq:Lmst}
\end{equation}
provide a quantitative measure of the degree of mass segregation: the larger $\Lmst$ the more concentrated are the massive stars compared to the
reference sample; the associated standard deviation $\Delta\Lmst$ quantifies the significance of the result. To make this method work and comparable
also with observational data all calculations are carried out in two dimensions, i.e. on a projection of the set of vertices.

We have chosen the number of random reference samples, $k$, such that a fraction $p$ of the entire population of size $N$ is covered independent of
the sample size $n$:
\begin{equation}
  p = 1 - \left( \frac{ N - n } { N } \right)^k \quad \Rightarrow \quad k = \lceil{ \frac{ \ln( 1 - p ) }{ \ln( 1 - n/N ) } }\rceil \,,
\end{equation}
where $\lceil{x}\rceil$ denotes the ceiling function\footnote{The ceiling function $\lceil{x}\rceil$ gives the lowest integer larger than or equal to
  $x$.} of $x$. The threshold has been set to $p = 0.99$.

Our method $\Mmstg$ involves two crucial modifications of $\Mmstl$ that make it computationally much more effective and boost its sensitivity. First,
unlike \citet{2009MNRAS.395.1449A} we do not calculate the separations between all possible pairs of stars but determine the MST (in two dimensions)
in a three-step procedure: first we use a 2D Delaunay triangulation \citep[from the software package GEOMPACK:][]{1991AdvEng.13..325J} to construct a
useful graph of stellar positions projected onto a plane, then we sort the edges of the triangles in ascending order, and finally adapt Kruskal's
algorithm \citep{1956PAMS....7...48K} with an efficient union-find-algorithm to construct the MST.

A Delaunay triangulation has the very useful property that the MST construction (in any dimension but for Euclidean distance) is its sub-graph
\citep{1985cgai.book.....P}. The implementation of \citet{1991AdvEng.13..325J} involves a computational effort
\begin{equation}
  t_{\mathrm{DT}} \sim \mathcal{O}(\left|E\right| \cdot \log \left|V\right|) \,,
\end{equation}
similar to the edge sorting by the quick-sort algorithm,
\begin{equation}
  t_{\mathrm{QS}} \sim \mathcal{O}(\left|E\right| \cdot \log \left|E\right|) \,,
\end{equation}
where $\left|E\right|$ is the number of edges and $\left|V\right|$ the number of vertices. With two algorithmic ``tricks'' in the union-find algorithm
(``union by rank'' and ``path compression'') we reduce its run-time to
\begin{equation}
  t_{\mathrm{UF}} \sim \mathcal{O}(\left|E\right| \cdot \log^* \left|V\right|) \,,
\end{equation}
where
\begin{equation}
  \log^*(n) = \min \Bigl\{s\in\mathbb{N} \mid \underbrace{\log\bigl(\log(\ldots\log(n)\ldots)\bigr)}_{s~\text{times}} \le 1 \Bigr\}
\end{equation}
and thus rather constant though in principle unlimited \citep{1979JACM...26..690T}.
So the total computational effort is dominated by the Delaunay triangulation and sorting of edges and thus scales as
\begin{equation}
  t_{\mathrm{tot}} \sim \mathcal{O}\bigl(\left|V\right| \cdot \log(\left|V\right|)\bigr) = \mathcal{O}\bigl(\left|E\right| \cdot
  \log(\left|E\right|)\bigr) \,,
\end{equation}
with $\mathcal{O}(|V|) = \mathcal{O}(|E|)$. This is a significant improvement over the cost of the brute-force method,
$\mathcal{O}(\left|E\right|^2)$.

Second, we do not use directly the \emph{sum} of the edges $\lmst$ as a measure yet their \emph{geometric mean} $\gmst$,
\begin{equation}
  \gmst = \bigg( \prod_{i=1}^n e_i \bigg)^{1/n} = \exp{ \left[ \frac{1}{n} \sum_{i=1}^n \ln{e_i} \right] } \,,
\end{equation}
and its associated geometric standard deviation $\Delta\gmst$,
\begin{equation}
  \Delta\gmst = \exp \left( \sqrt{ \sum_{i=1}^n ( \ln e_i - \ln \gmst )^2 \over n } \right) \,,
\end{equation}
where $e_i$ are the $n$ MST edges.
Analogous to $\Lmst$, we obtain the new measure $\Gmst$ via a proper normalisation: 
\begin{equation}
  \Gmst       = \frac{ \gref }{ \gmass } \,, \quad
  \Delta\Gmst = \Delta\gref \,.
  \label{eq:Gmst}
\end{equation}
Note that the usage of a geometric mean involves a multiplicative standard deviation $\Delta\Gmst$, i.e. the upper and lower 1$\sigma$ intervals are
given by $\Gmst \cdot (\Delta\Gmst)^{\pm1}$.

The geometrical mean has two important properties that turn out to be very useful for our purpose of using the MST as a measure of mass segregation:
\begin{enumerate}
\item The $n$-th root implicitly involves an intermediate-pass that damps contributions from extreme edge lengths very effectively (i.e. it gives a
  lower weight to very short or very long edges). Hence the mean edge length of a compact configuration of even few stars will not be affected much by
  an ``outlier''.

  We demonstrate this property via a simple case of $n-1$ edges with length $l = e_0$ and one very short ``outlier'' representing a tight binary with
  $l = e_1 = \epsilon e_0$, where $\epsilon \ll 1$ ($e_1$ and $\epsilon$ could be also very large here in principle). Then the geometric mean would
  yield
  \begin{equation}
    <l> = \left( e_0^{n-1} e_1 \right)^{1/n} = e_0^n \epsilon^{1/n} \,.
  \end{equation}
  In any practical case (i.e. realistic star clusters models, observational data) the relevant binary separation will be at most 3 orders of magnitude
  smaller than the mean separation of single stars (or binary centre-of-mass).\footnote{This estimate is based on the following reference values: the
    mean stellar separation in the Orion Nebula Cluster is about $2.5\,\mathrm{pc} / 4000^{1/3} \approx 0.15\,\mathrm{pc} \approx 30000\,\mathrm{AU}$;
    the HST has a resolution of $\sim$50\,AU at an assumed distance of 400\,pc
    \citep{2007A&A...474..515M,2007MNRAS.376.1109J,2009A&A...497..195K}. Thus, binaries with separations $\lesssim 10^{-3}$ would remain unresolved.}
  So with $\epsilon = 10^{-3}$ we obtain even for a very small sub-sample of $n = 5$ stars a mean edge length $<l> \approx 1/4 e_0$. This demonstrates
  the very effective damping of extreme values by the geometric mean.
\item The product of edges has the valuable property that common edges in the two samples of massive and reference stars, $\gmass$ and $\gref$, are
  cancelled out by normalisation, Eq.~\eqref{eq:Gmst}. So it is only the disjoint set of edge lengths that determines $\Gmst$ and hence the degree of
  mass segregation. Compared to $\Mmstl$ our scheme $\Mmstg$ is thus much more robust when applied to stellar systems with a high binary fraction.
\end{enumerate}

\subsection{Primordial mass-segregation}

\label{sec:methods:segregation}

\begin{figure}
  \centering
  \includegraphics[height=0.9\linewidth,angle=-90]{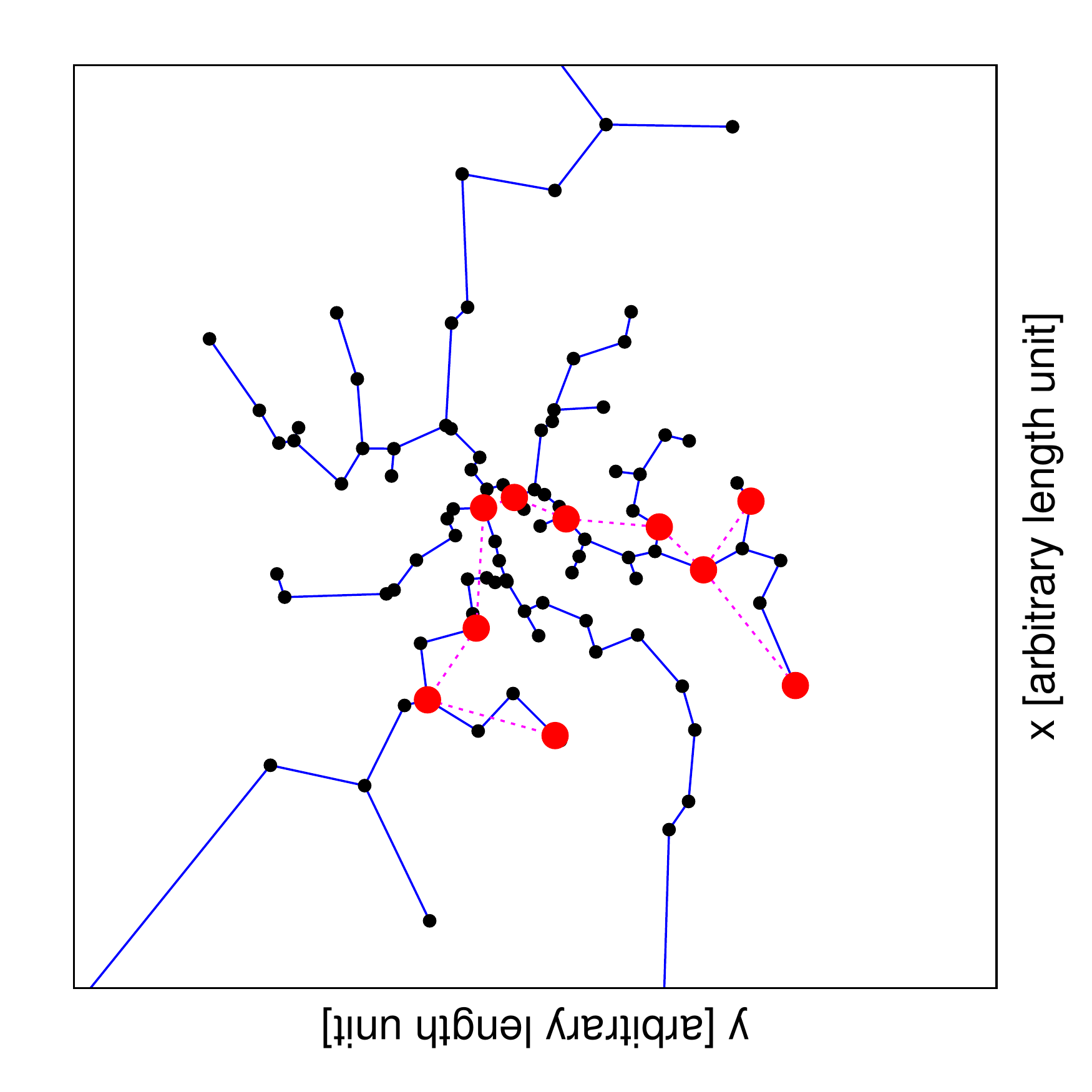}
  \includegraphics[height=0.9\linewidth,angle=-90]{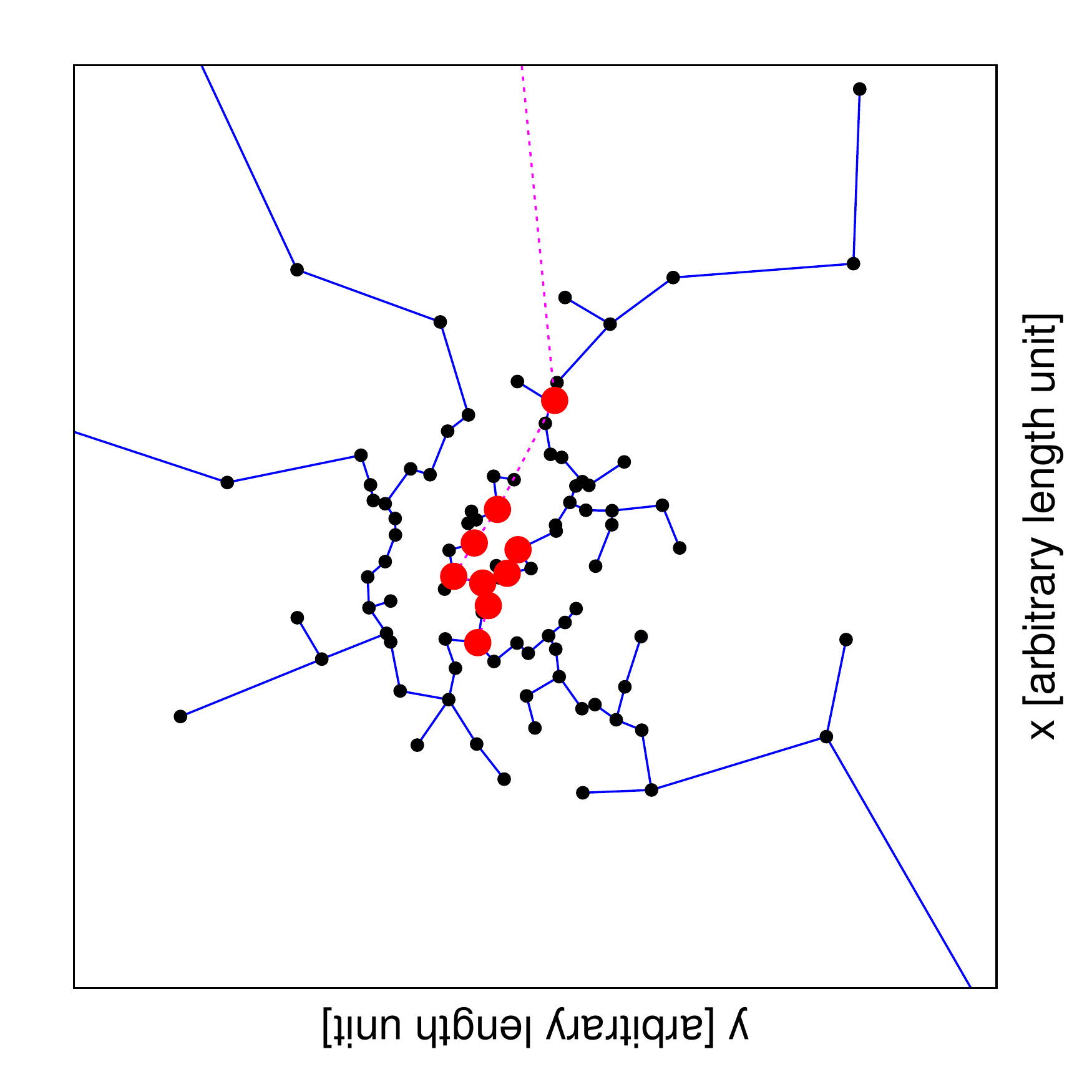}
  \caption{Visualisation of the two-dimensional minimum spanning tree (MST) for numerical cluster models with different initial mass segregation
    parameters $S$ (see Section~\ref{sec:methods:segregation}). The solid blue lines (connecting the small black dots and large red dots) represent
    the MST of the entire population. The red dashed lines (connecting the large red dots) represent the MST of the ten most massive
    particles. \emph{Top:} $S = 0.1$. \emph{Bottom:} $S = 0.9$.}
  \label{fig:mst__subr}
\end{figure}

A thorough test of our method $\Mmstg$ requires the generation of predefined mass segregated stellar systems. For this purpose we found the method of
\citet{2008MNRAS.385.1673S} to be very useful. The strength of the method is the generation of a well defined degree of mass segregation that is
controlled via one single parameter, $S$, and the creation of dynamically consistent models by modelling local potentials and velocities in a
quasi-equilibrium state. The so-called \emph{index of mass segregation} covers the range $S \in [ 0, 1)$, where $S=0$ corresponds to an entirely
unsegregated state, while the upper limit $S=1$ marks the maximum possible segregation. In Fig.~\ref{fig:mst__subr} we show two initially mass
segregated models of 100 particles with $S = 0.1$ and $S= 0.9$ and their corresponding MST, respectively.

In summary, the authors find that a mean inter-particle potential of the form
\begin{equation}
  \mmean{U^{ij}} = 2(1 - S)^2 \mmean{U_\mathrm{tot}}\, \frac{m_i\,m_j}{M_\mathrm{c}^2} \, \left(
    \frac{M_\mathrm{sub}^i\,M_\mathrm{sub}^j}{M_\mathrm{c}^2} \right)^{-S} \,,
  \label{eq:mean_potential}
\end{equation}
using ordered subsets of stars,
\begin{equation}
  M_\mathrm{sub}^i \equiv \sum_{j=1}^i \, m_j\;,\;\; m_1 \geq m_2 \geq\,...\,\geq m_N\;,
\end{equation}
provides one of the simplest forms that characterise locally consistent mass segregation. This equation provides a constraint on the distribution
function from which one can construct a cluster with the desired degree of mass segregation $S$ by adding one-by-one the individual stars from a
mass-ordered set. The underlying distribution function
\begin{equation}
  n(r) \propto r^2 \left(r_\mathrm{p}^2(M_\mathrm{sub}^i) + r^2 \right)^{-5/2}
\end{equation}
with
\begin{equation}
  r_\mathrm{p}(M_\mathrm{sub}^i) = \frac{3\pi}{32}\, \frac{GM_\mathrm{c}^2}{|\mmean{U_\mathrm{tot}}|}\, \frac{1}{1-S} \left(
    \frac{M_\mathrm{sub}^i}{M_\mathrm{c}} \right)^{2S} \,,
  \label{eq:rp}
\end{equation}
provides a good estimate of the real distribution function of a Plummer sphere with mass segregation index $S$.

The authors provide a numerical C-code \textsc{plumix} for generating the cluster according to the algorithm described in their paper on the AIfA web
page: http://www.astro.uni-bonn.de.

%

\section{Results}

\label{sec:results}

In this section we provide examples showing the much better performance of $\Mmstg$ than $\Mmstl$. However, we will also compare with the more
traditional method $\Mmfd$ introduced in Section~\ref{sec:introduction}.

\subsection{Special configurations}

\label{sec:results:special}

\begin{figure}
  \centering
  \includegraphics[height=0.9\linewidth,angle=-90]{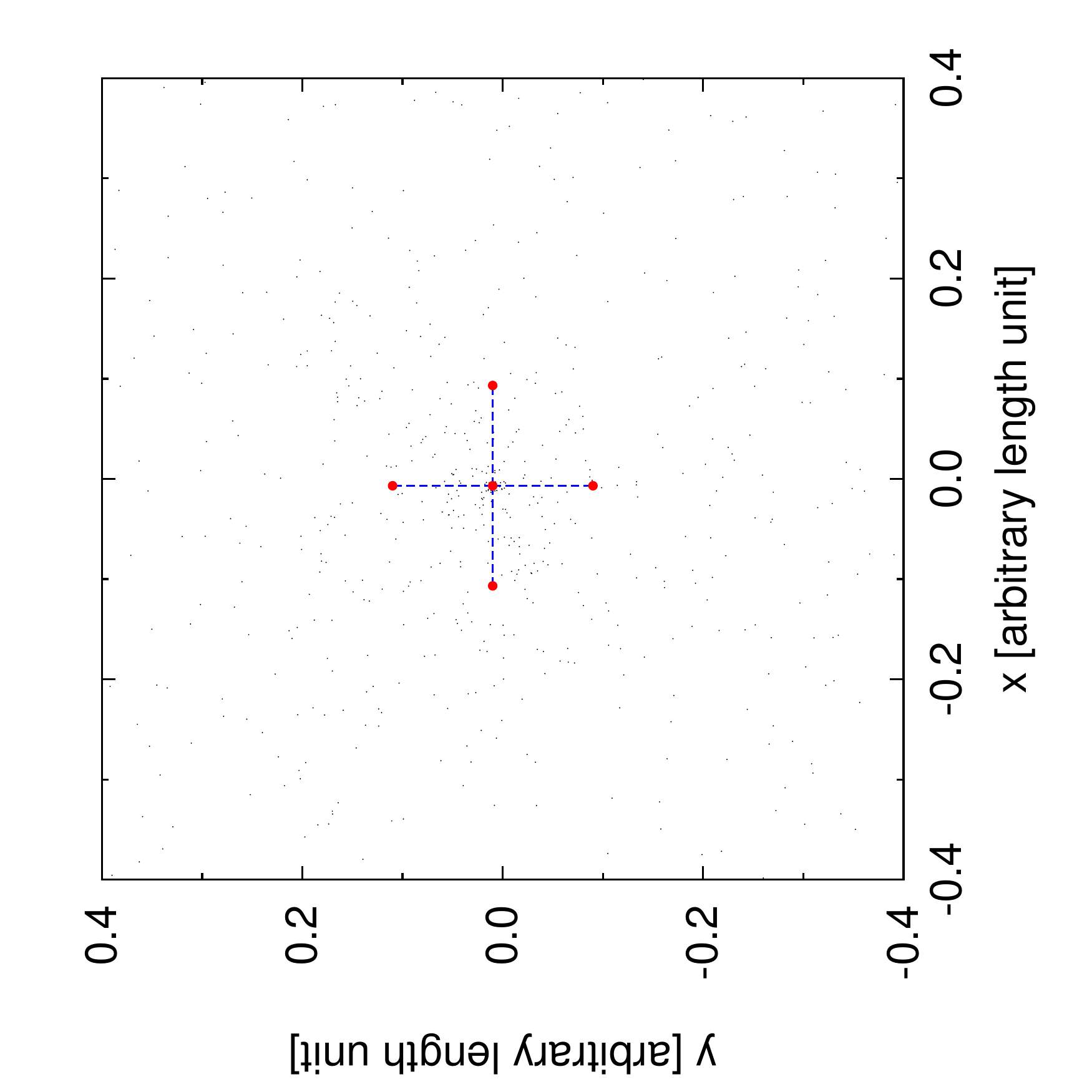}
  \includegraphics[height=0.9\linewidth,angle=-90]{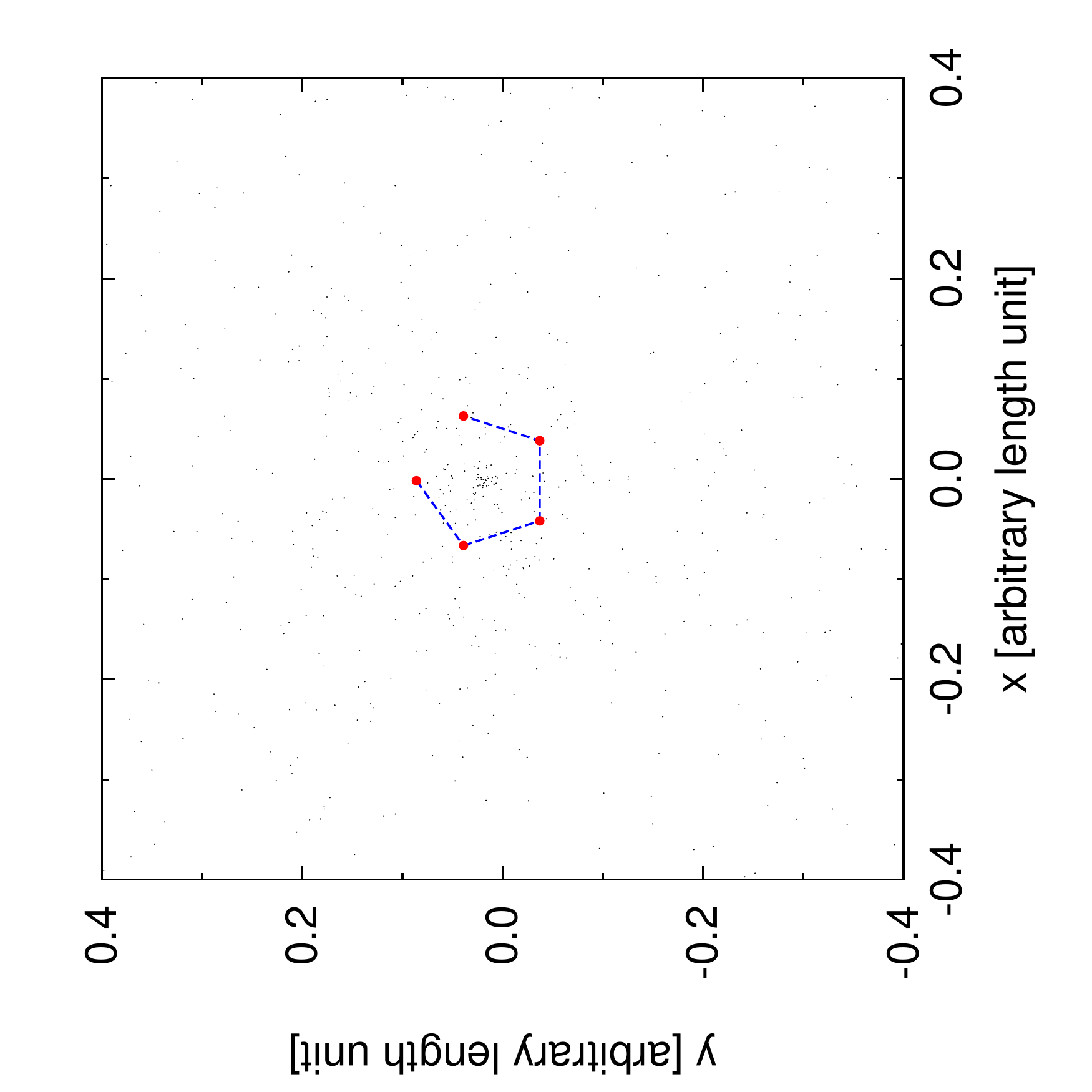}
  \includegraphics[height=0.9\linewidth,angle=-90]{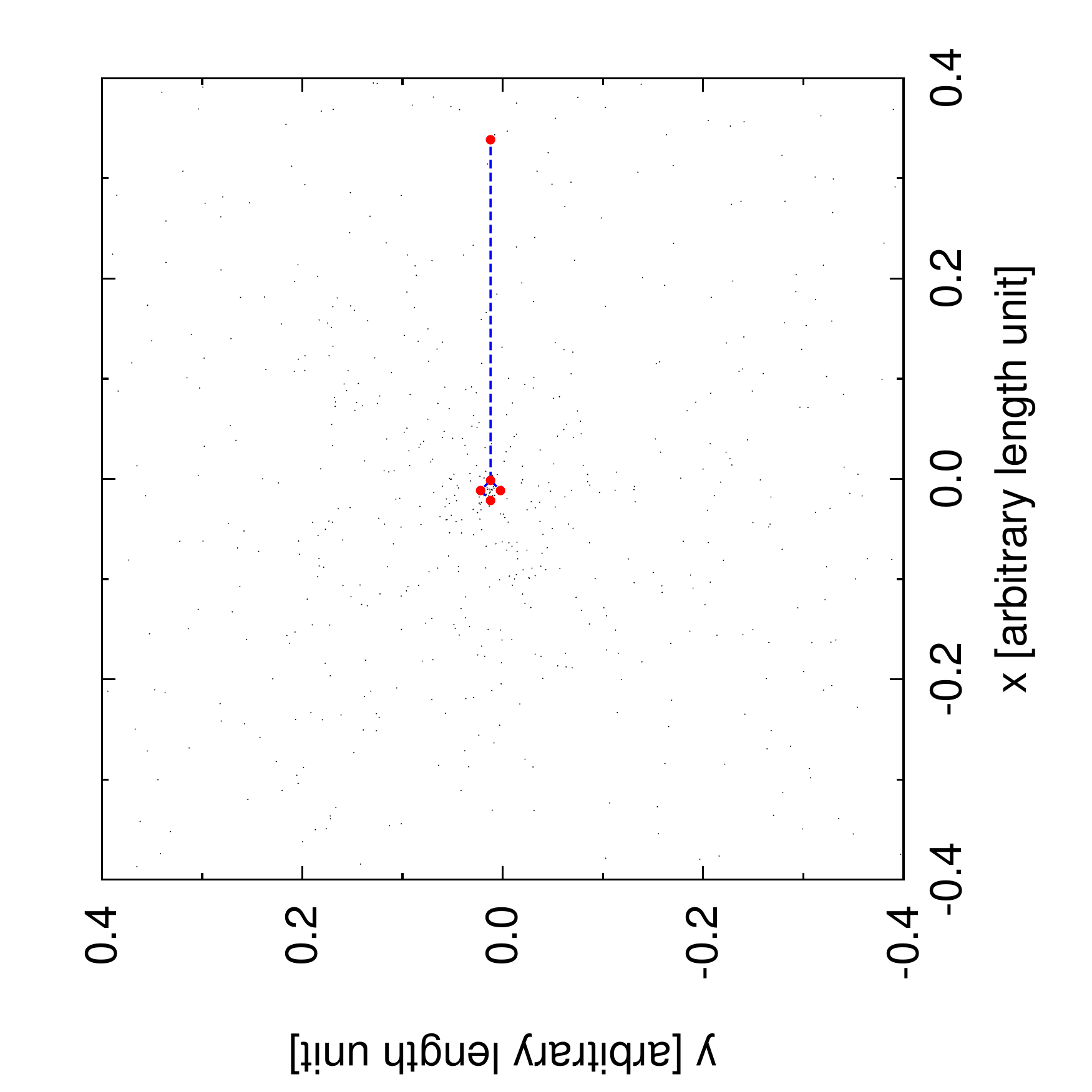}
  \caption{Three artificial configurations of massive stars with identical $\lmst$ in a model star cluster of 1k members. \textit{Top:} ``cross'',
    \textit{Middle:} ``ring'', \textit{Bottom:} ``clump''.}
  \label{fig:mst__5_most_massive}
\end{figure}

First, we will demonstrate the power of $\Mmstg$ for some simple setups of artificial mass segregation as shown in Fig.~\ref{fig:mst__5_most_massive}.

The idea to improve $\Mmstl$ and develop $\Mmstg$ was in fact motivated by the goal to find a measure that reflects the optical impression of a lower
or higher degree of mass segregation. The plots in Fig.~\ref{fig:mst__5_most_massive} depict three different artificial configurations of the five
most massive stars in a star cluster of 1k members, designated ``cross'', ``ring'', and ``clump'', from top to bottom. All these configurations are
characterised by identical $\lmst$, i.e. according to $\Mmstl$ they represent the same degree of mass segregation. However, from an observer's point
of view the cross appears to show a lower degree of mass segregation than the ring (i.e. the latter appears more compact), while the clump would
usually be interpreted as a highly segregated system with a peculiar outlier.

\begin{figure}
  \centering
  \includegraphics[height=0.9\linewidth,angle=-90]{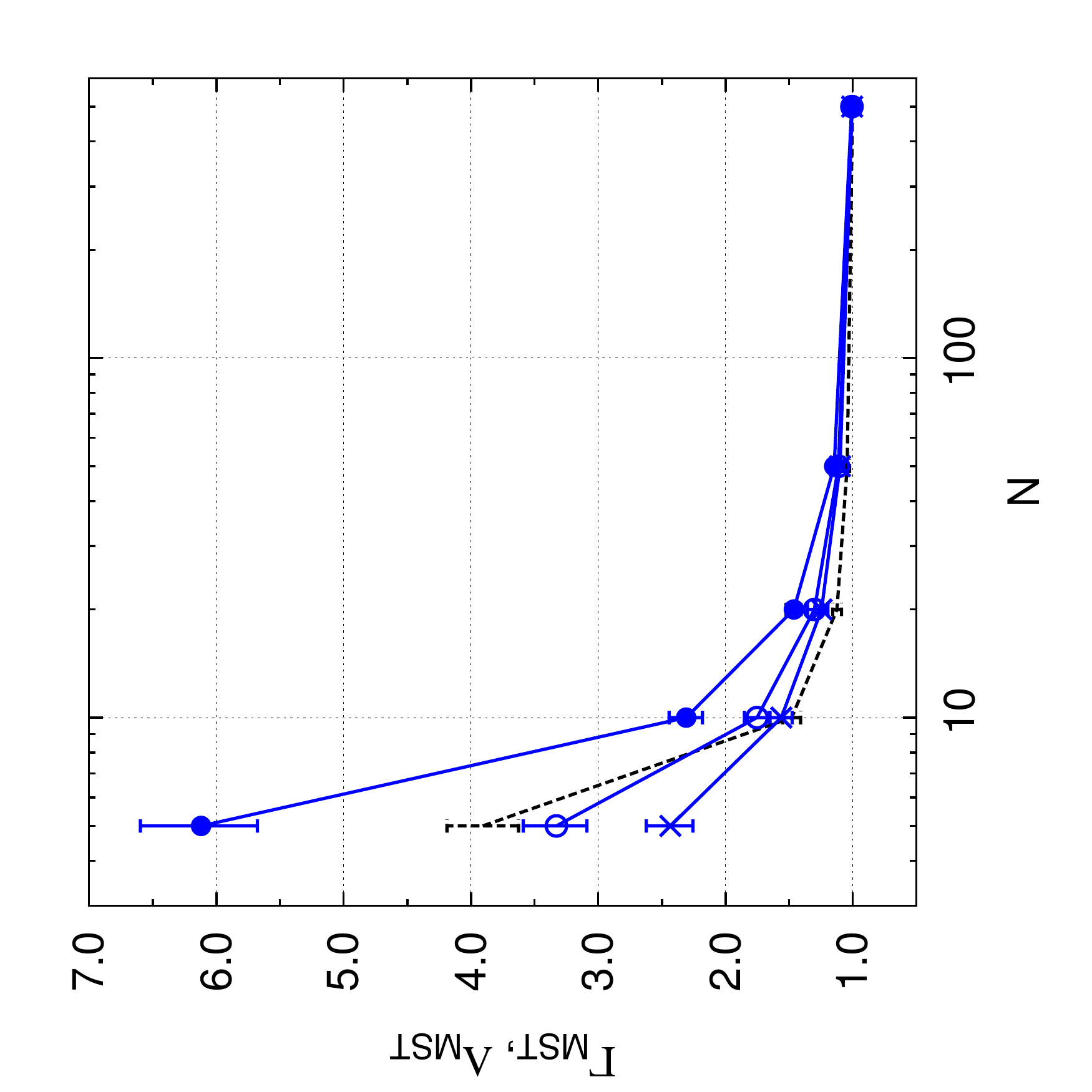}
  \caption{Comparison of $\Gmst$ (blue solid lines) and $\Lmst$ (black dashed line) of three artificial configurations of massive stars in a model
    star cluster. From top to bottom, the solid lines with their symbols represent the artificial configuration ``clump'' (closed circles), ``ring''
    (open circles), and ``cross'' (crosses).}
  \label{fig:mst__5_most_massive__table}
\end{figure}

Translated into a consistent algorithm we aim at measuring the compactness of a stellar system by assigning a higher weight to the dominant
configuration of stars and hence to overcome the ``degeneracy'' of $\Mmstl$. As already discussed in Section~\ref{sec:methods:mst} this is mainly
achieved by damping the contribution from ``outliers'', i.e. single edges with extreme lengths $e_i$ compared to the median of all edges, via the
geometrical mean. Fig.~\ref{fig:mst__5_most_massive__table} demonstrates the effect of $\Mmstg$. While by construction $\Lmst$ (black dashed line) is
identical for all three configurations, $\Gmst$ depends strongly on the degree of concentration of the dominant sample of stars. Using the five most
massive stars for the calculation of $\Gmst$ the highly concentrated ``clump'' shows a roughly two times higher degree of mass segregation than the
other two configurations. Also, it's significance (i.e. standard deviation in relation to the mean) is about 1.5 times higher than in the case of
$\Lmst$.

Note that though only the five most massive stars form a centrally concentrated configuration, i.e. are mass-segregated, one obtains a signature of
mass segregation for a sample of up to 20 most massive stars. In particular, $\Gmst$ of the ten most massive stars does even depend on the geometry of
the configuration.

\subsection{Initially mass-segregated clusters}

\label{sec:results:initial}

Using the method of \citet{2008MNRAS.385.1673S} to create initially mass-segregated star clusters (see Section~\ref{sec:methods:segregation}) we have
generated a set of low- to high-mass star clusters with 100, 1k, 10k, and 100k stars, respectively. Their underlying mass function from
\citet{2001MNRAS.322..231K} in the range $0.08 - 150\,\Msun$ results in an average stellar mass of $\sim$$0.6\,\Msun$. So our clusters span the mass
range from $60\,\Msun$ to $60\mathrm{k}\,\Msun$ which basically represents the entire observed cluster population in our Galaxy. All clusters have
been set up with initial degrees of mass segregation $S = \{0.1, 0.2, 0.3, 0.5, 0.9\}$, each with ten configurations with different random seeds to
reduce statistical uncertainties. However, in the following plots we show rescaled error bars that represent the statistical uncertainties of one
single cluster.

\begin{figure*}
  \centering
  \includegraphics[height=0.45\linewidth,angle=-90]{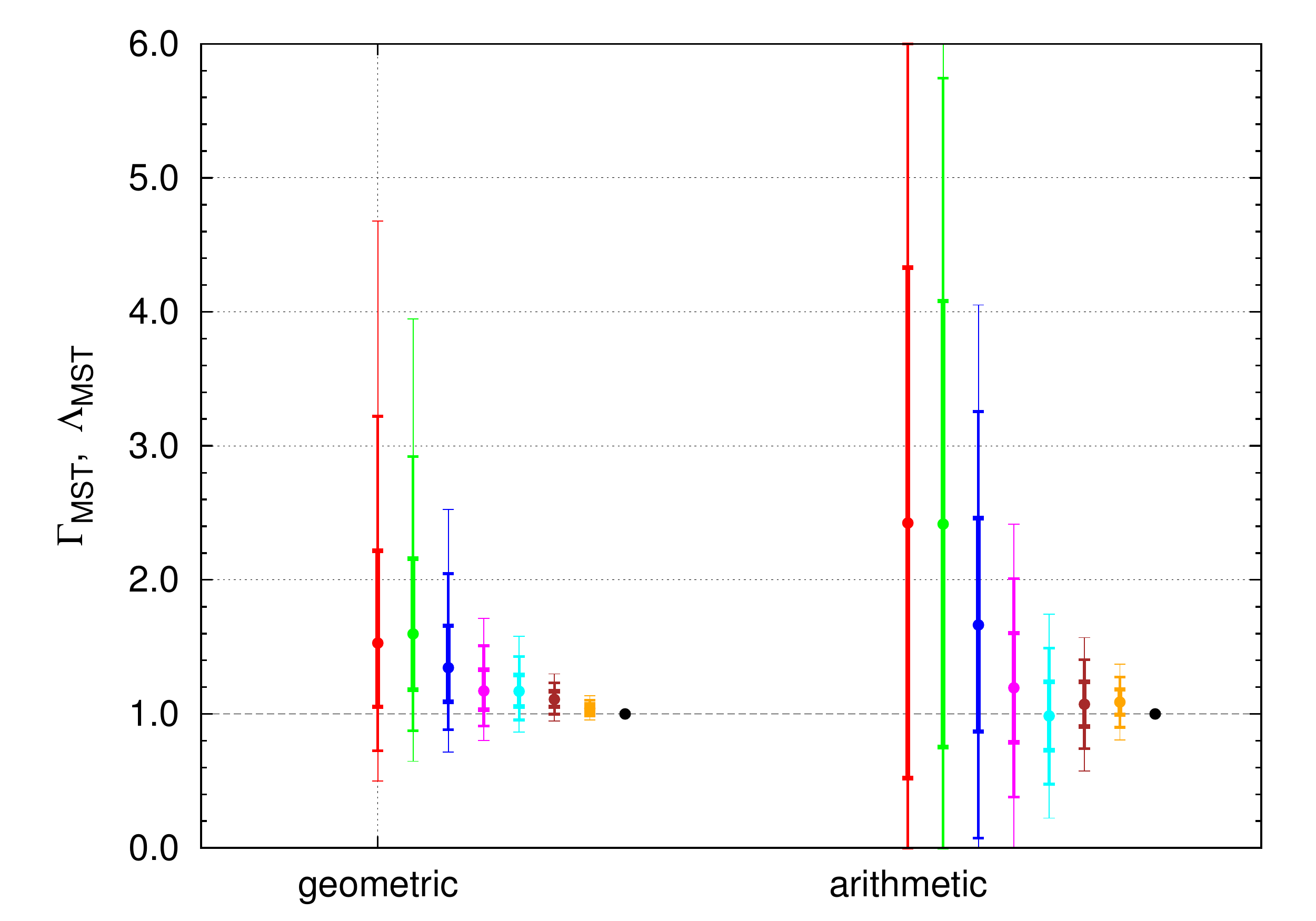}
  \includegraphics[height=0.45\linewidth,angle=-90]{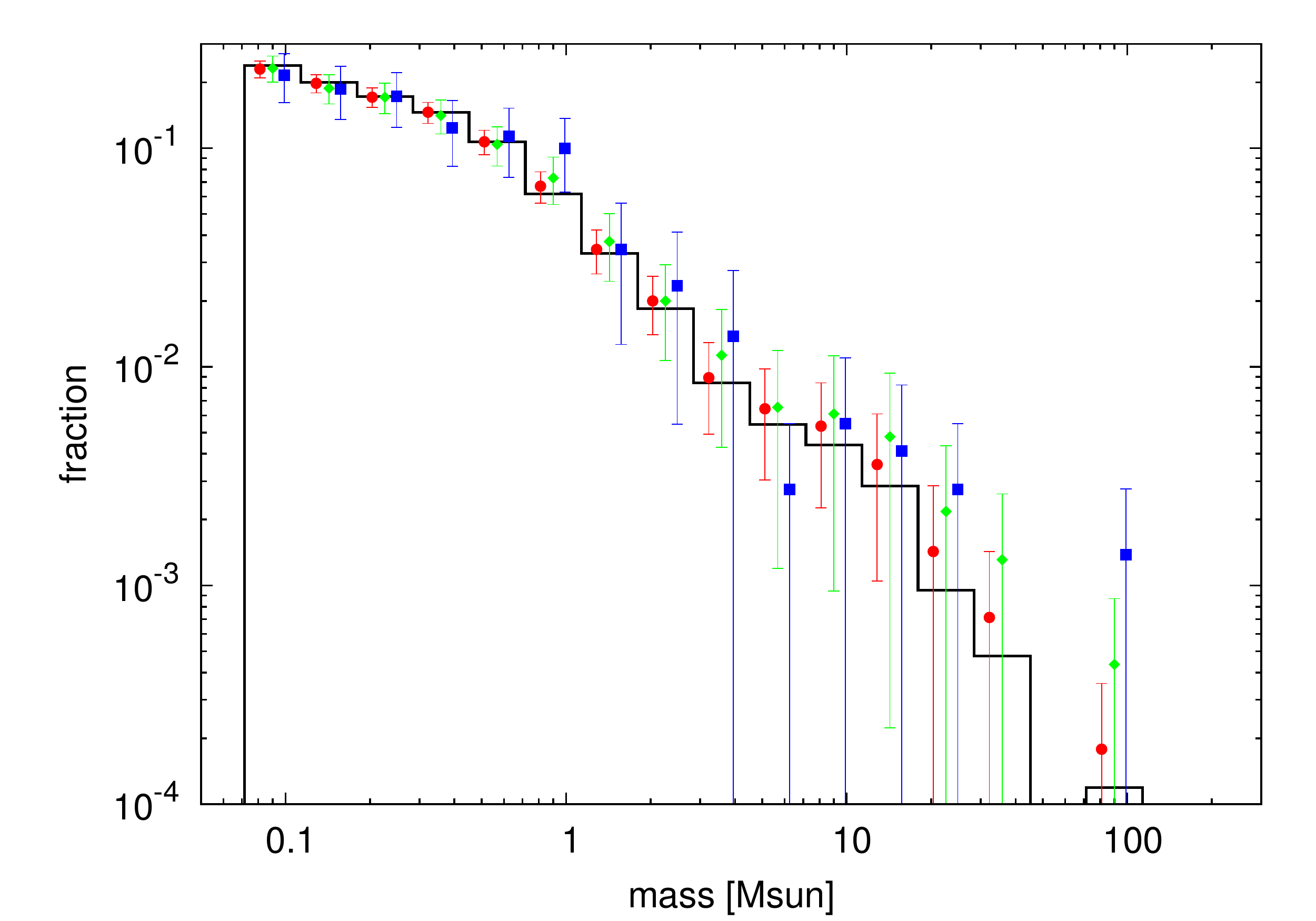}
  \includegraphics[height=0.45\linewidth,angle=-90]{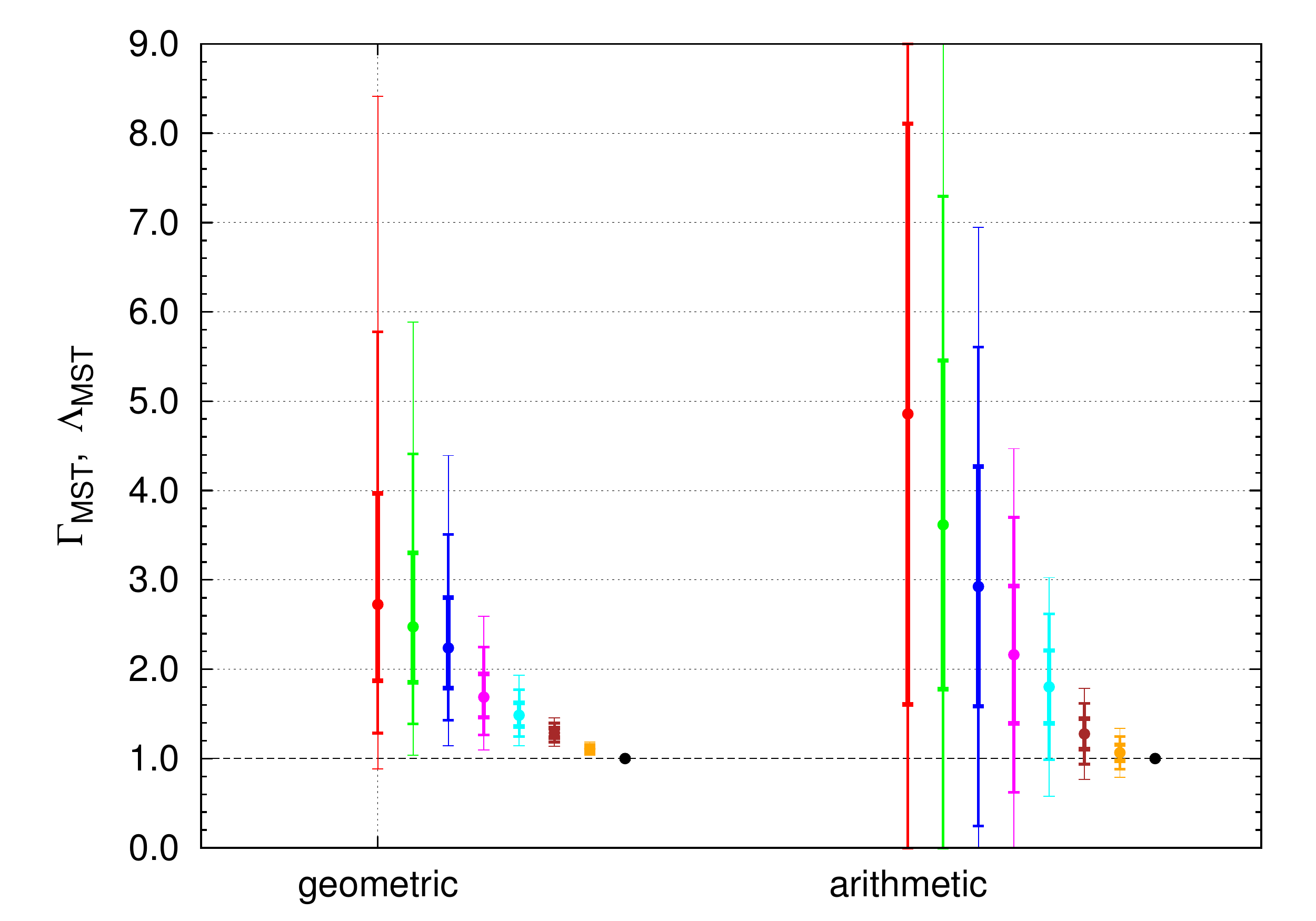}
  \includegraphics[height=0.45\linewidth,angle=-90]{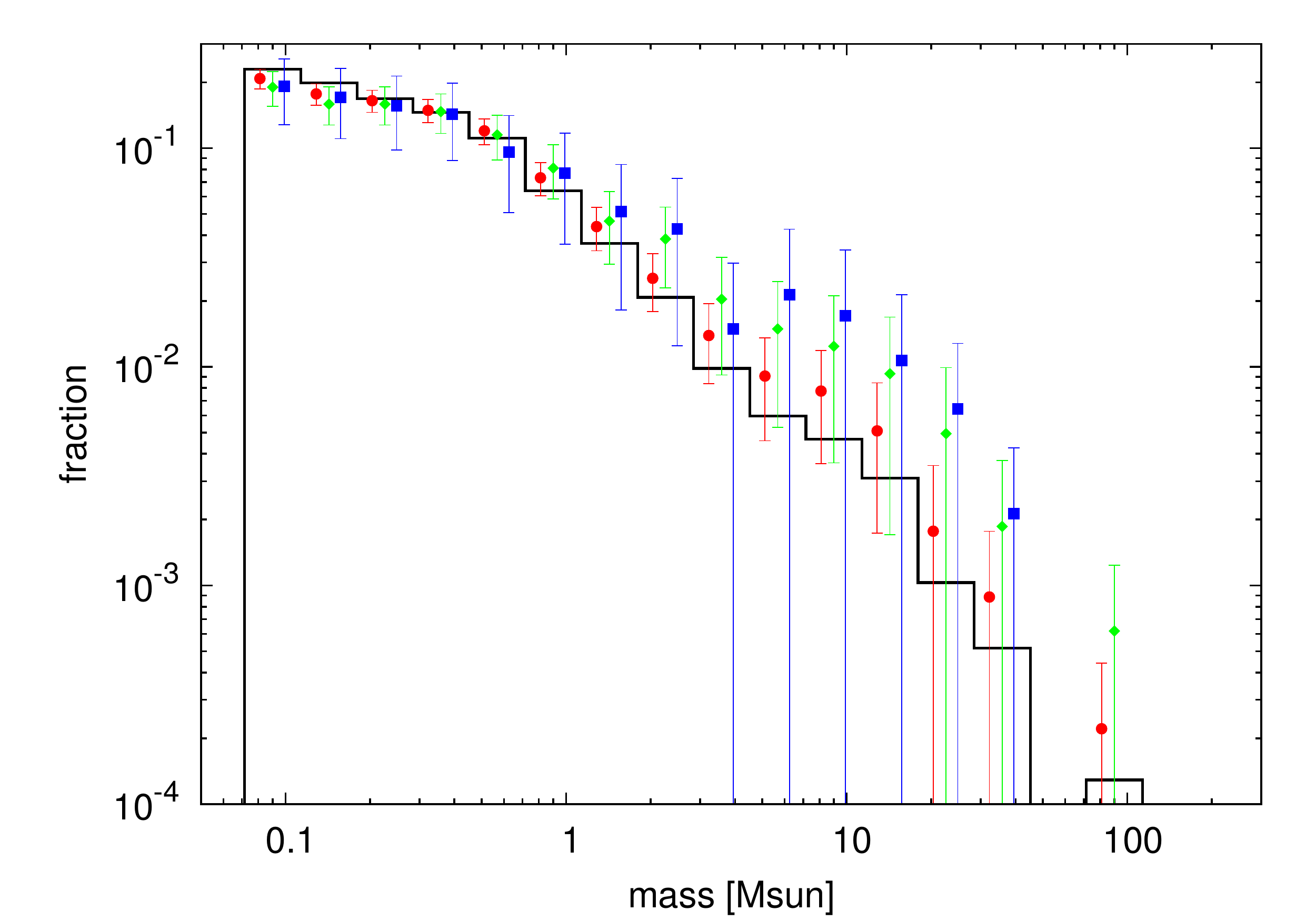}
  \includegraphics[height=0.45\linewidth,angle=-90]{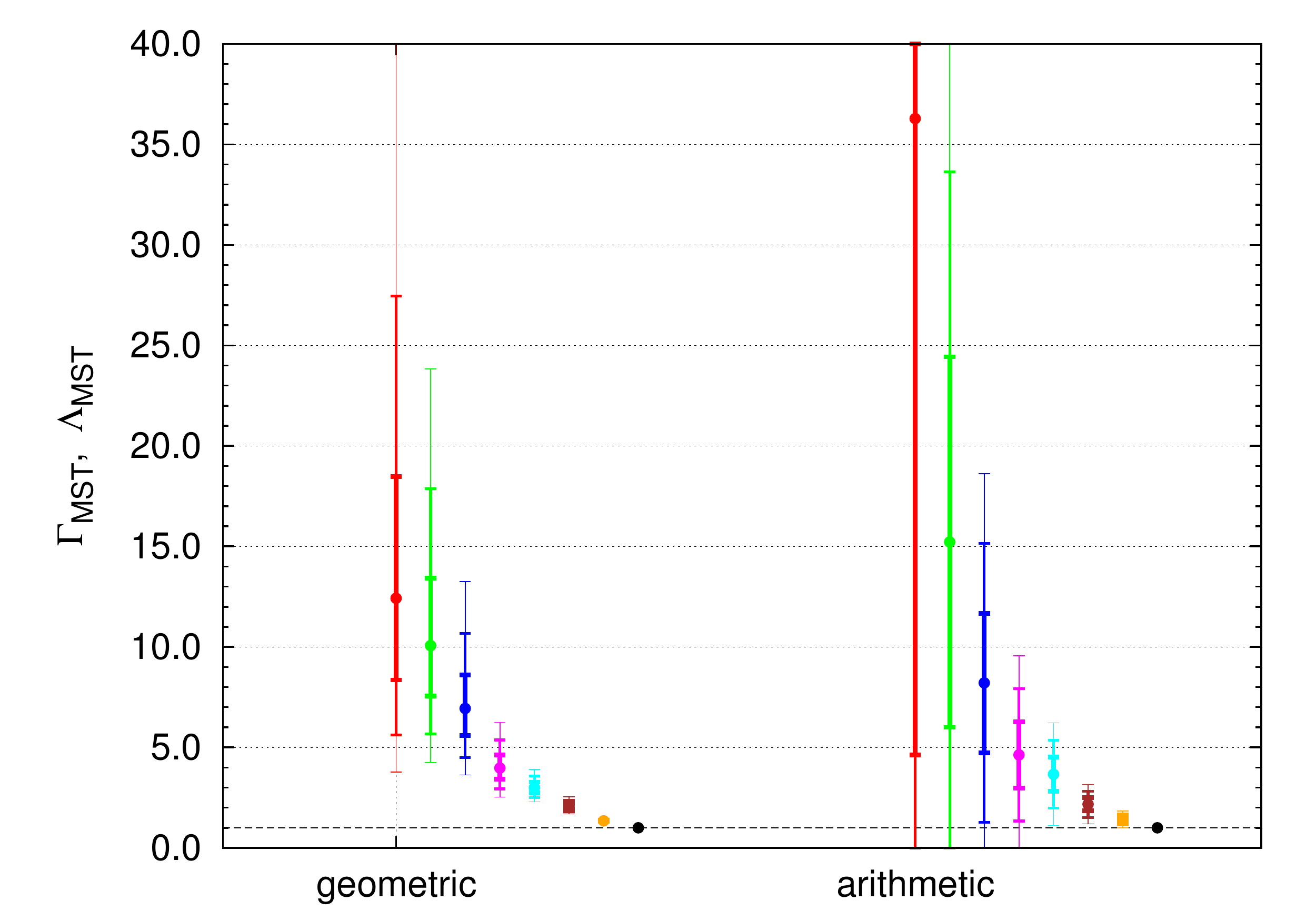}
  \includegraphics[height=0.45\linewidth,angle=-90]{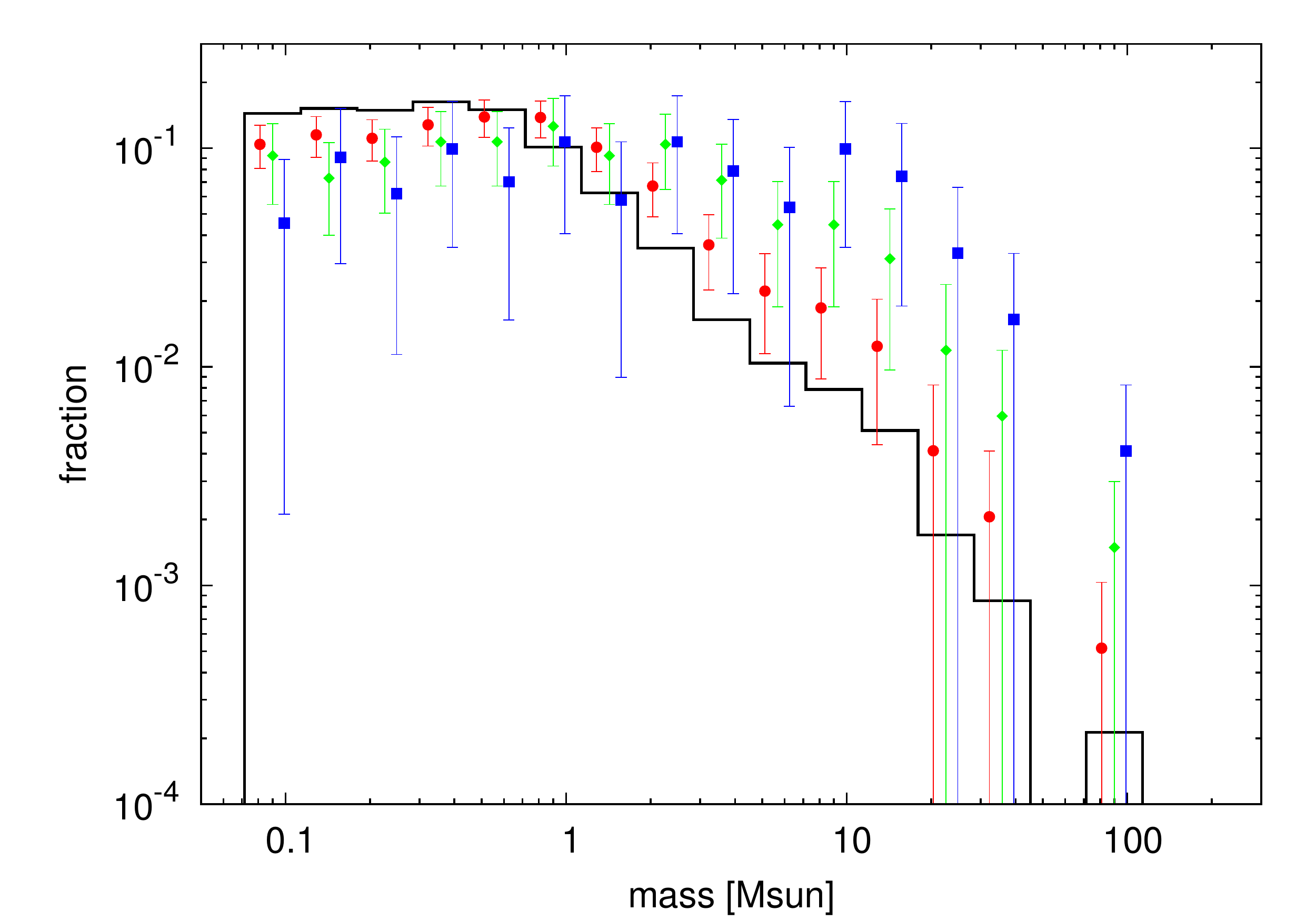}
  \caption{Diagnostics of initially mass segregated star clusters with 1k members following the prescription of \citet{2008MNRAS.385.1673S}. From top
    to bottom the degree of mass segregation, $S$, equals 0.1, 0.3, and 0.9 (see text for more details). On the left-hand side we compare $\Gmst$ and
    $\Lmst$ for the 5, 10, 20, 50, 100, 200, 500, 1000 most massive stars. The error bars and line thickness mark the $1\sigma$, $2\sigma$, and
    $3\sigma$ uncertainties. On the right-hand side we plot the corresponding mass function of the entire cluster population (solid black line), and
    the population within one (red tics), one-half (green crosses), and one-forth (blue dots) half-mass radius. The error bars mark the $1\sigma$
    uncertainties.}
  \label{fig:mass_segregation__initial__small}
\end{figure*}

\begin{figure*}
  \centering
  \includegraphics[height=0.45\linewidth,angle=-90]{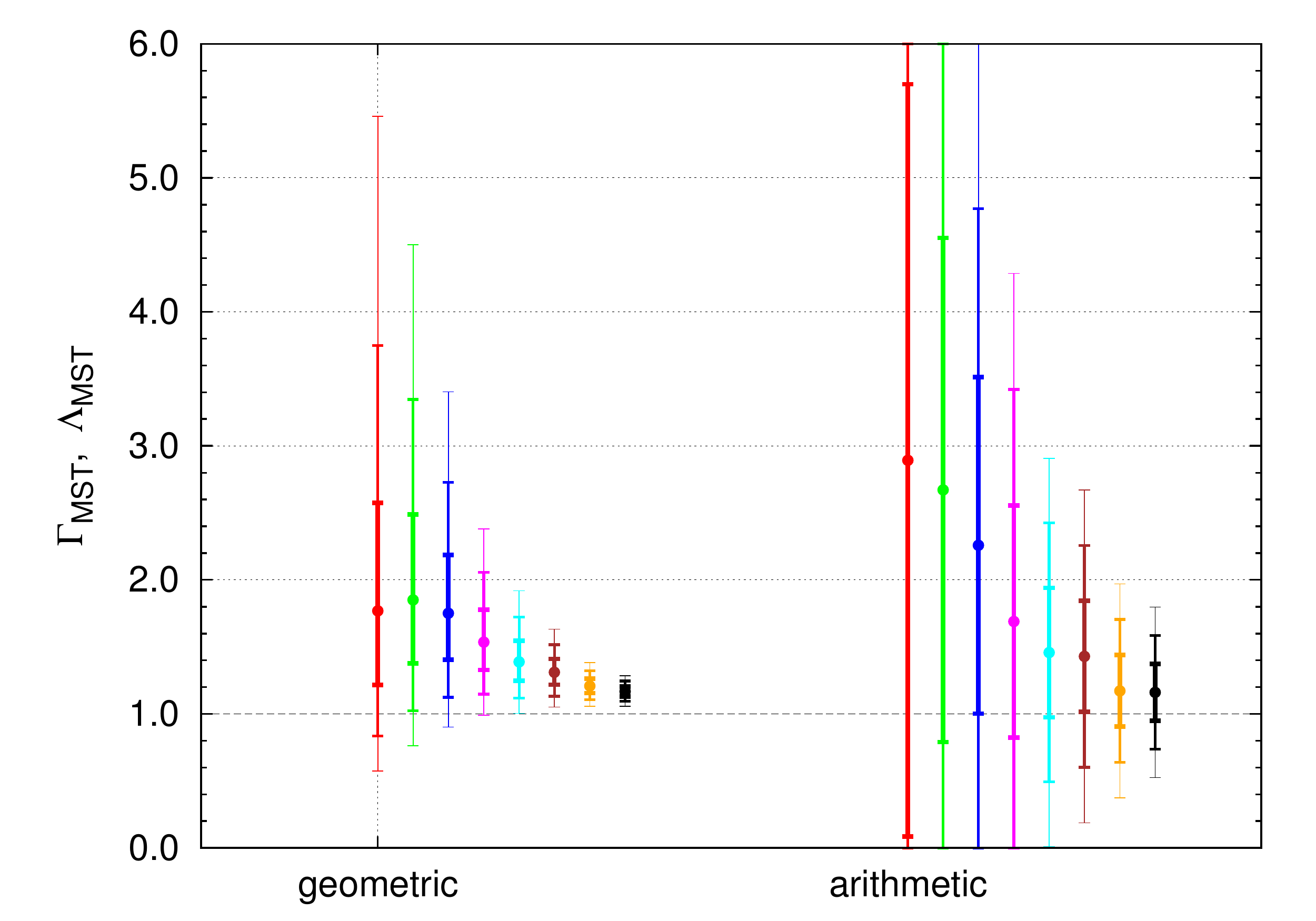}
  \includegraphics[height=0.45\linewidth,angle=-90]{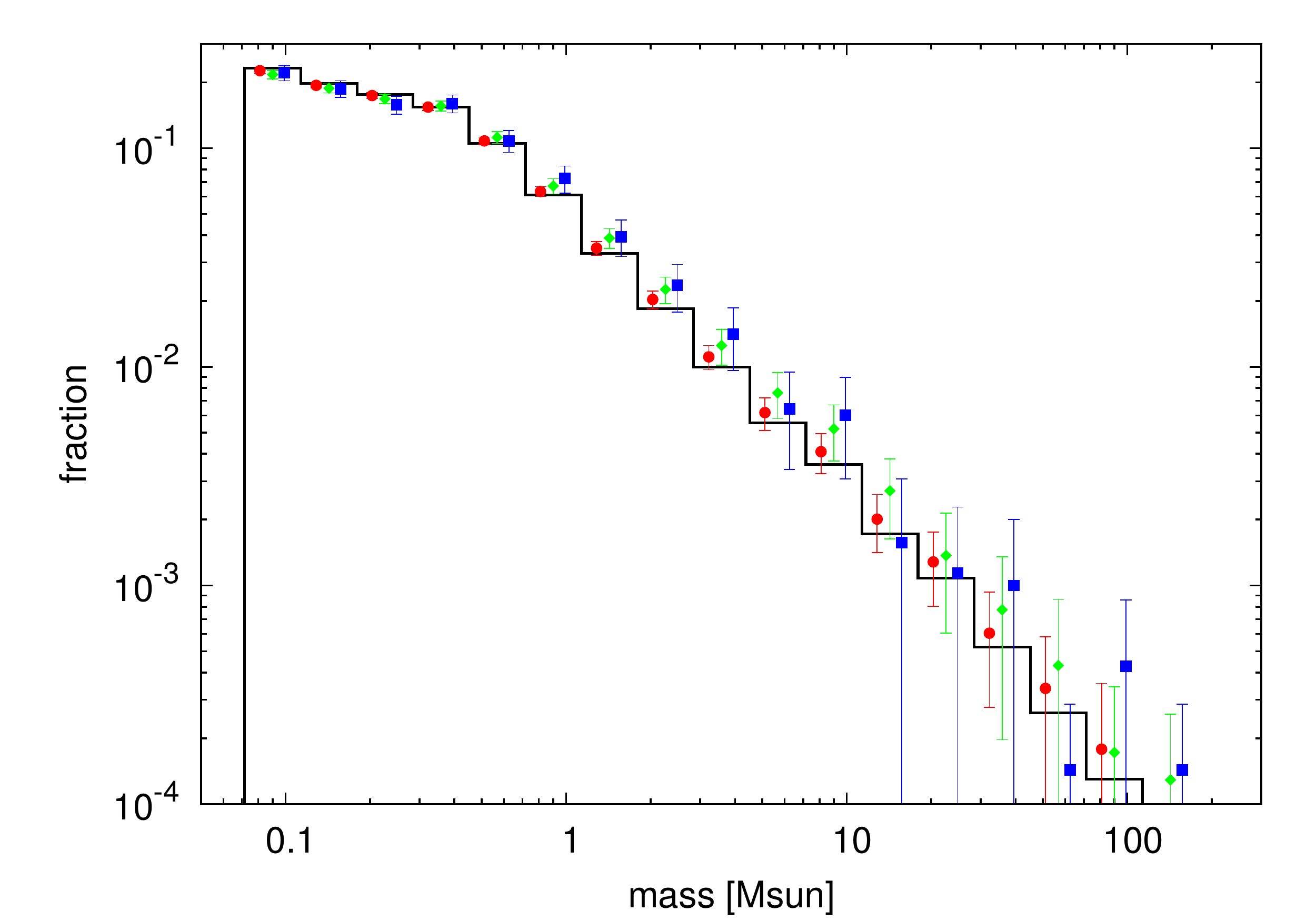}
  \includegraphics[height=0.45\linewidth,angle=-90]{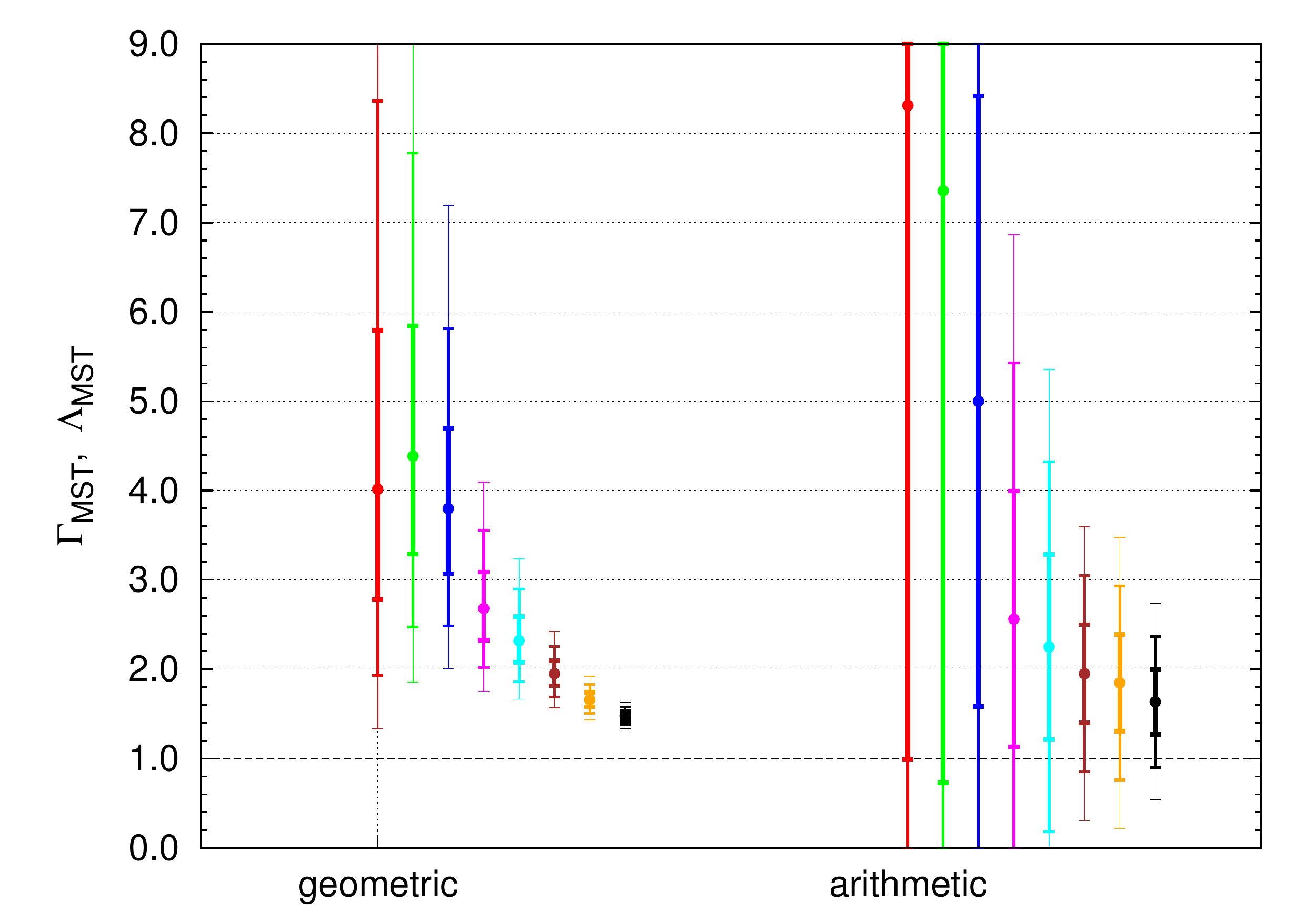}
  \includegraphics[height=0.45\linewidth,angle=-90]{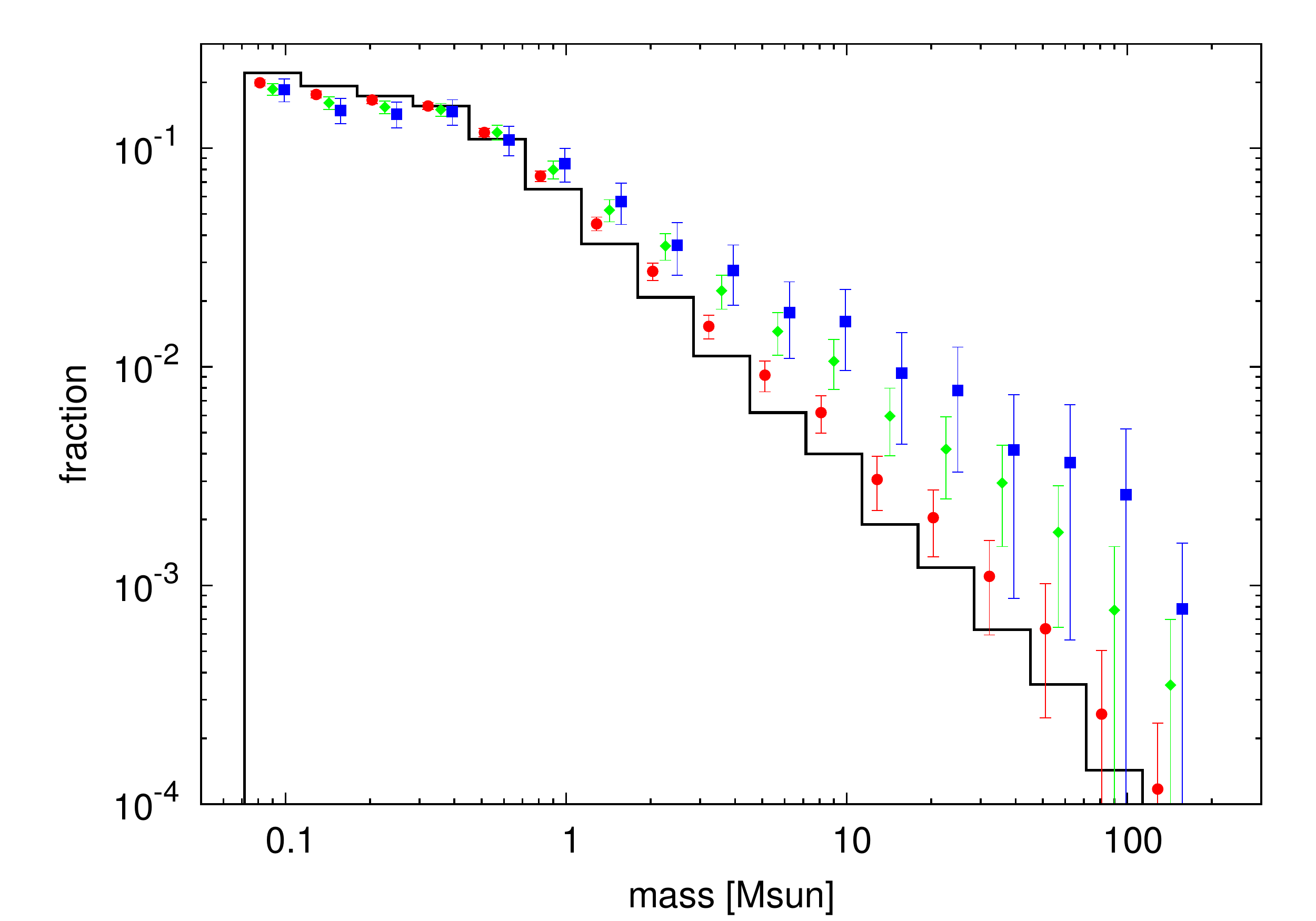}
  \caption{Diagnostics of initially mass segregated star clusters with 10k members following the prescription of \citet{2008MNRAS.385.1673S}. From top
    to bottom the degree of mass segregation, $S$, equals 0.1 and 0.3 (see text for more details). Details of the plots are given in the caption of
    Fig.~\ref{fig:mass_segregation__initial__small}.}
  \label{fig:mass_segregation__initial__massive}
\end{figure*}

In Figs.~\ref{fig:mass_segregation__initial__small} and \ref{fig:mass_segregation__initial__massive} we show a selection of the most relevant
results. On the left-hand side we compare $\Gmst$ (``geometric'') and $\Lmst$ (``arithmetic'') for the 5 (red), 10 (green), 20 (blue), 50 (magenta),
100 (cyan), 200 (brown), 500 (orange), and 1000 (black) most massive stars. The error bars and line thickness mark the $1\sigma$, $2\sigma$, and
$3\sigma$ uncertainties. The horizontal dotted lines mark integer values, the horizontal dashed line represents the unsegregated state $\Gmst = \Lmst
= 1$.  Star clusters of intermediate masses $\sim$1000\,$\Msun$ are well observed in our local neighbourhood. These objects are represented by our
simulations with 1k particles. From Fig.~\ref{fig:mass_segregation__initial__small} we find that our measure of mass segregation, $\Gmst$, detects a
very low degree of mass segregation, $S = 0.1$, just above the $1\sigma$-level, nearly independent of the number of most massive stars considered. The
significance improves drastically for an intermediate degree of $S = 0.3$ to at least $3\sigma$ and even up to $4\sigma$ for the 20 to 100 most
massive stars. Finally, for very strong mass segregation with $S = 0.9$ we obtain a very clear signature of more than $5\sigma$ for any number of
(i.e. at least five) most massive stars.

These results are very much better than using $\Lmst$ which \emph{never} approaches a significance of $3\sigma$ and - in particular - for $S = 0.3$
provides only a very weak $1\sigma$ significance compared to $3-4\sigma$ in the case of $\Gmst$.

On the right-hand side we plot the corresponding mass functions for comparison with the traditional method $\Mmfd$ to detect mass segregation. Here
the mass function has been calculated in annuli with a radius $r = \{ 1/4 r_{\mathrm{h}}, 1/2 r_{\mathrm{h}}, r_{\mathrm{h}}, r_{\mathrm{c}} \}$,
where $r_{\mathrm{h}}$ is the half-mass and $r_{\mathrm{c}}$ the total radius of the star cluster. It is clearly demonstrated that only very strong
mass segregation with $S \approx 0.9$ becomes evident via this method. In this case it is the mass range of stars around $10\,\Msun$ that provides the
strongest signature.

In contrast, our new measure $\Gmst$ is much more effective. In particular, the 10 to 20 most massive stars of a 1k star cluster usually provide the
clearest signature of mass segregation.

Qualitatively, we find the same results for more massive clusters of 10k stars in Fig.~\ref{fig:mass_segregation__initial__massive}: $\Mmstg$ provides
the best measure of mass segregation by far. However, there are some important quantitative differences. First, a much lower degree of mass
segregation can be detected for more massive clusters, e.g. for 10k stars with $S = 0.1$ a significance of $3\sigma$ is reached. Second, for a similar
significance mass segregation becomes most evident for a larger number of most massive stars, e.g. the 20 most massive stars for a 10k system with $S
= 0.3$ compared to the 10 most massive for a 1k system with $S = 0.9$.

Both effects are in agreement with our expectations that in a more populous star cluster i) the same degree of mass segregation $S$ will result in a
higher relative concentration of the same number of most massive stars, i.e. $\Gmst$ becomes larger, and ii) the absolute number of stars that show
the same relative concentration is larger, i.e. $\Delta\Gmst$ becomes lower.

The interplay of these two properties explains the existence and dependence of the local maximum of the significance of $\Gmst$: it peaks at the
maximum ratio of $\Gmst$ and $\Delta\Gmst$ and thus increases for larger cluster masses and lower degrees of mass segregation towards larger numbers
of most massive stars.

From our entire set of initially mass-segregated clusters we find that a sample size of 10 to 20 most massive stars generally provides the clearest
signature of mass segregation.

Note that from Figs.~\ref{fig:mass_segregation__initial__small} and \ref{fig:mass_segregation__initial__massive} we conclude that using $\Mmfd$ the
best estimate of mass segregation in star clusters from 1k to 10k members would be for masses at $\sim$10\,$\Msun$ in an annulus of radius $r \approx
1/4 r_{\mathrm{h}}$.

\subsection{Mass-segregation in the ONC}

\label{sec:results:onc}

We recall that one of our goals was to develop a method that \emph{equally} well applies to data from simulations and observations and thus
demonstrate the excellent performance of $\Mmstg$ on observational data of the $\sim$1\,Myr old Orion Nebula Cluster (ONC) obtained by
\citet{1997AJ....113.1733H}. The sample contains 929 stars with mass estimates and so provides a robust test of the algorithm.

\begin{figure}
  \centering
  \includegraphics[height=0.95\linewidth,angle=-90]{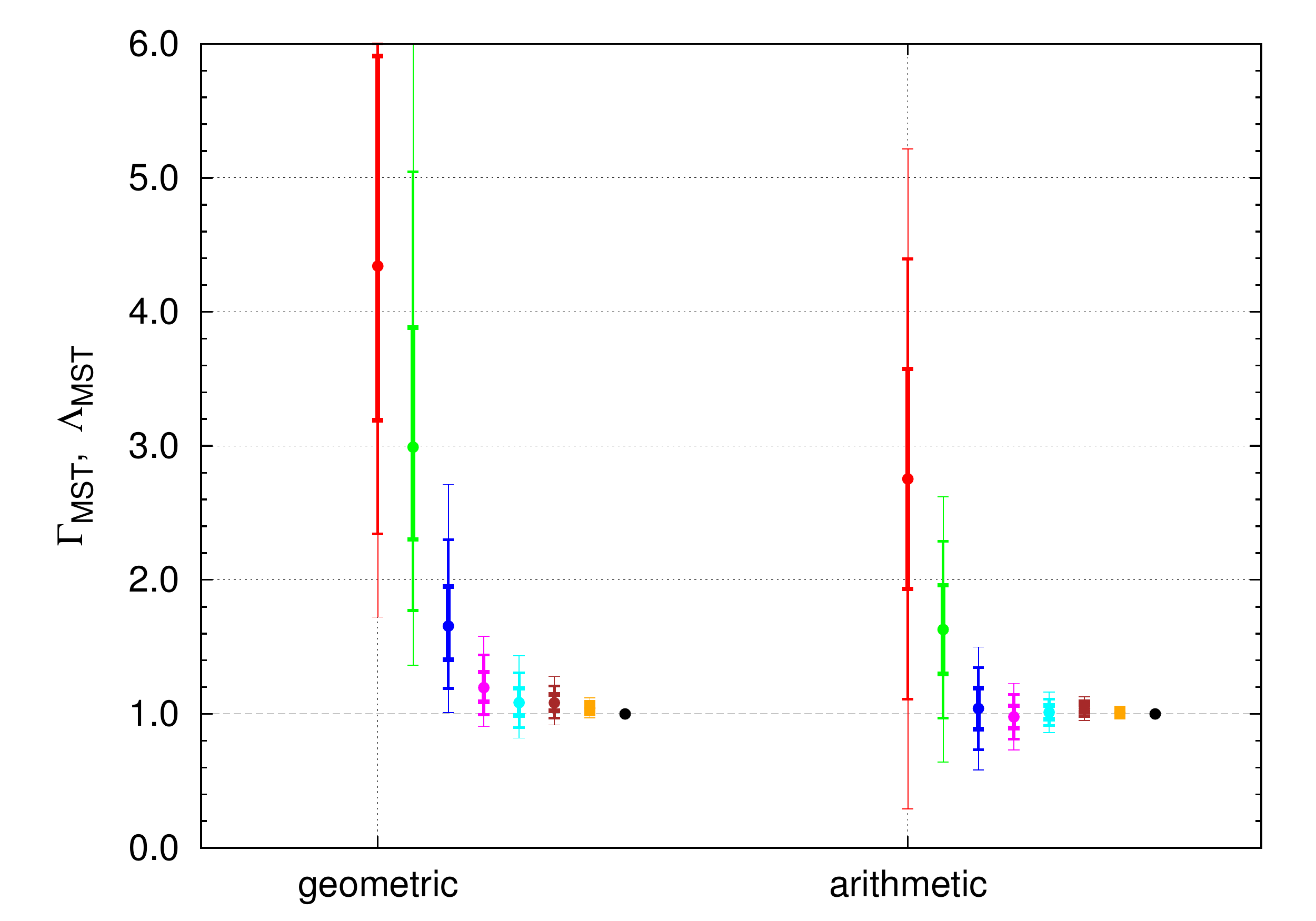}
  \includegraphics[height=0.95\linewidth,angle=-90]{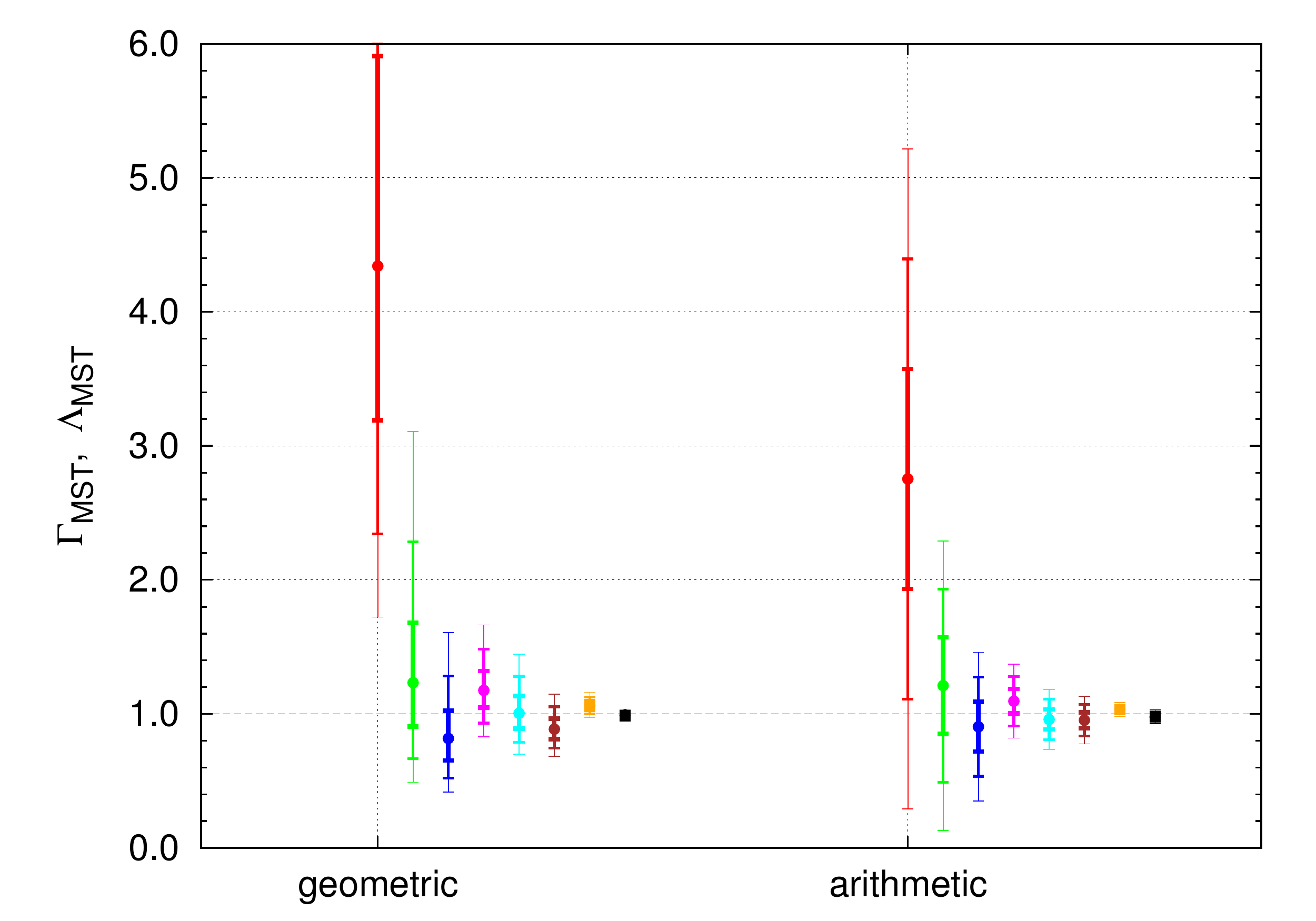}
  \caption{Diagnostics of mass segregation in the young Orion Nebula Cluster (ONC) using observational data from
    \citet{1997AJ....113.1733H}. \emph{Top:} $\Gmst$ and $\Lmst$ for the 5, 10, 20, 50, 100, 200, 500 most massive, and all 929 stars (left to
    right). \emph{Bottom:} $\Gmst$ and $\Lmst$ for the 5, 6 to 10, 11 to 20, 21 to 50, 51 to 100, 101 to 200, 201 to 500, and 500 to 929 most massive
    stars (left to right). The error bars and line thickness mark the $1\sigma$, $2\sigma$, and $3\sigma$ uncertainties.}
  \label{fig:mass_segregation__onc}
\end{figure}

The resulting plot of $\Gmst$ for the 5, 10, 20, 50, 100, 200, 500 most massive, and all 929 stars in the upper part of
Fig.~\ref{fig:mass_segregation__onc} provides a clear signature of a significant concentration of the 20 most massive stars, much stronger than given
by $\Lmst$. Interestingly, our calculation of $\Lmst$ shows slightly stronger mass segregation than published by \citet{2009ApJ...700L..99A}. Compared
to the initially mass-segregated cluster models with 1k particles presented in Fig.~\ref{fig:mass_segregation__initial__small} the plot of $\Gmst$ for
the ONC shows a more complex distribution: there is a sharp drop between the 10 and 20 most massive stars. The significance of the first two data
points ($\gtrsim$$4\sigma$) corresponds to the model with $S = 0.9$, the $\lesssim$$3\sigma$ significance of the other resembles the much less
segregated models with $S \leq 0.3$.

In summary, i) mass segregation in the ONC at the current age $\sim$1\,Myr is much stronger than estimated before, and ii) it is clearly detectable
not only for the five most massive stars yet for the 20 most massive objects.

However, the particular features of the $\Gmst$ distribution require a more careful analysis. We do so by calculating $\Gmst$ of disjoint particle
groups, i.e. the most massive 5, 6 to 10, 11 to 20, etc. stars. The corresponding plot at the bottom of Fig.~\ref{fig:mass_segregation__onc} shows
that it is only the five most massive stars that are really strongly mass segregated (with a significance $>$$4\sigma$). Their configuration is so
dominant that adding the next 15 most massive stars - which are not segregated - still shows a strong signature at the $3\sigma$ level, similar to the
results in Section~\ref{sec:results:special}.

\citet{1997AJ....113.1733H} argues that her sample of combined photometric and spectroscopic data appears completely representative of the ONC,
showing in particular a uniform completeness with cluster radius. However, because observational data are usually biased by incompleteness this issue
is addressed in Appendix \ref{app:incompleteness}. We demonstrate that incompleteness has only a moderate effect on $\Mmstg$ and provide a simple
prescription for recovering the original signature of mass segregation via the completeness function.

We conclude that all but the five most massive ONC stars do \emph{not} show any signs of mass segregation. This result is in excellent agreement with
\citet[][Fig.~4]{2006ApJ...644..355H}. The five most massive stars' extraordinary degree of mass segregation resembles their peculiar tight
trapezium-like configuration \citep[e.g.][]{1994AJ....108.1382M,1998ApJ...492..540H}.

\subsection{Dynamical evolution of mass-segregation}

\label{sec:results:dynamical}

As a last application of our new method $\Mmstg$ we analyse again numerical data yet here we trace the dynamical evolution of a star cluster model
over time. The initial configuration is based on our ONC model as described in previous publications \citep[e.g.][]{2010A&A...509A..63O}. In short, it
is a single star cluster with an isothermal initial density distribution, a Maxwellian velocity distribution, the mass function of
\citet{2001MNRAS.322..231K} in the range $0.08-150\,\Msun$, and an initial size of 2.5\,pc. Note that we use a spherically symmetric model with a
smooth density distribution without any substructure. However, in contrast to our basic equilibrium model here our stellar system is initially
collapsing, starting from a virial ratio $Q = 0.1$. The simulations were carried out with $\nbody$\footnote{Note that we have used the GPU enabled
  version of $\nbody$ available at http://www.ast.cam.ac.uk/research/nbody.} \citep[][]{2003grav.book.....A} until a physical age of 5\,Myr,
corresponding to 13.5 N-body time units of the cluster.

\begin{figure}
  \centering
  \includegraphics[height=0.95\linewidth,angle=-90]{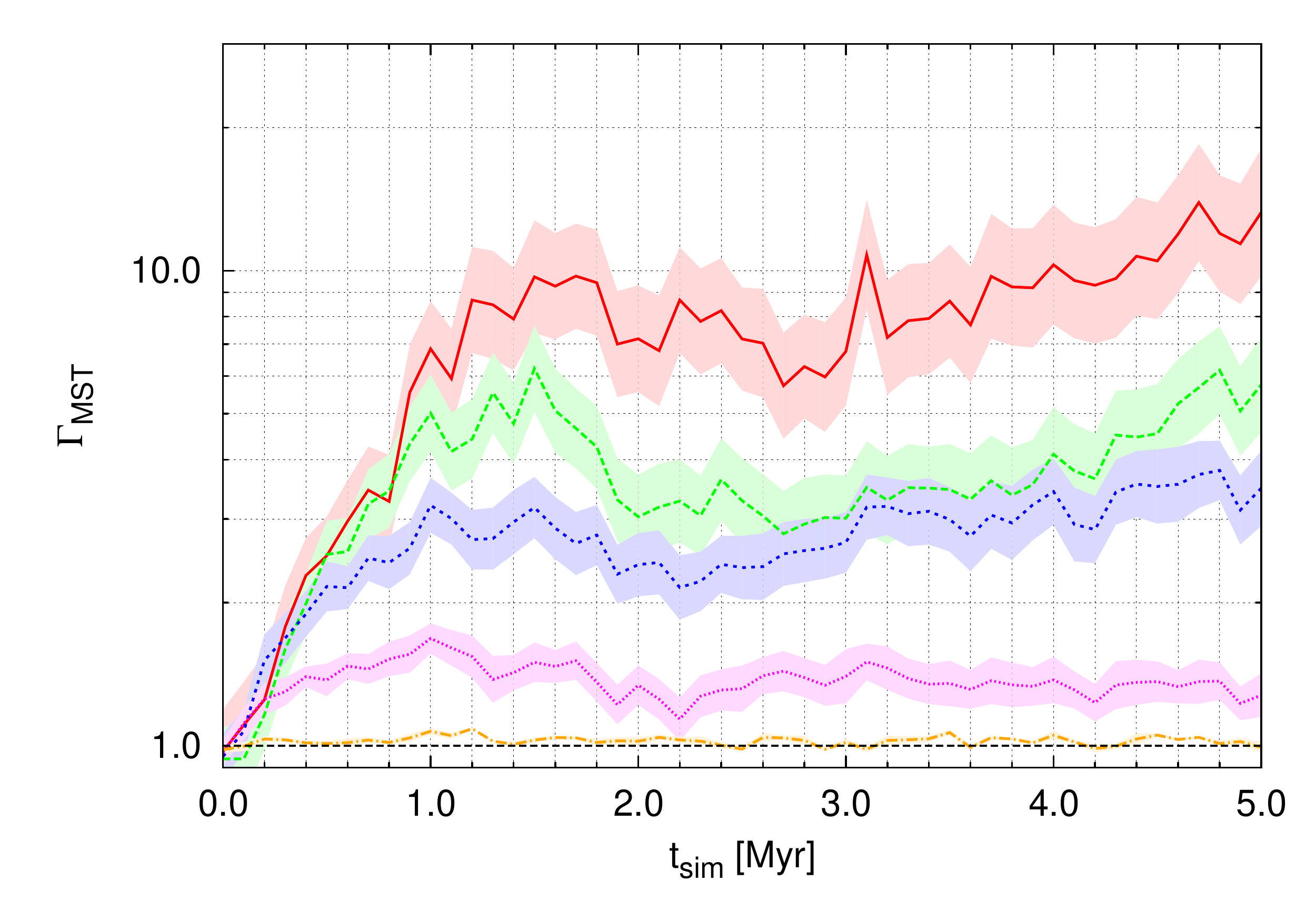}
  \includegraphics[height=0.95\linewidth,angle=-90]{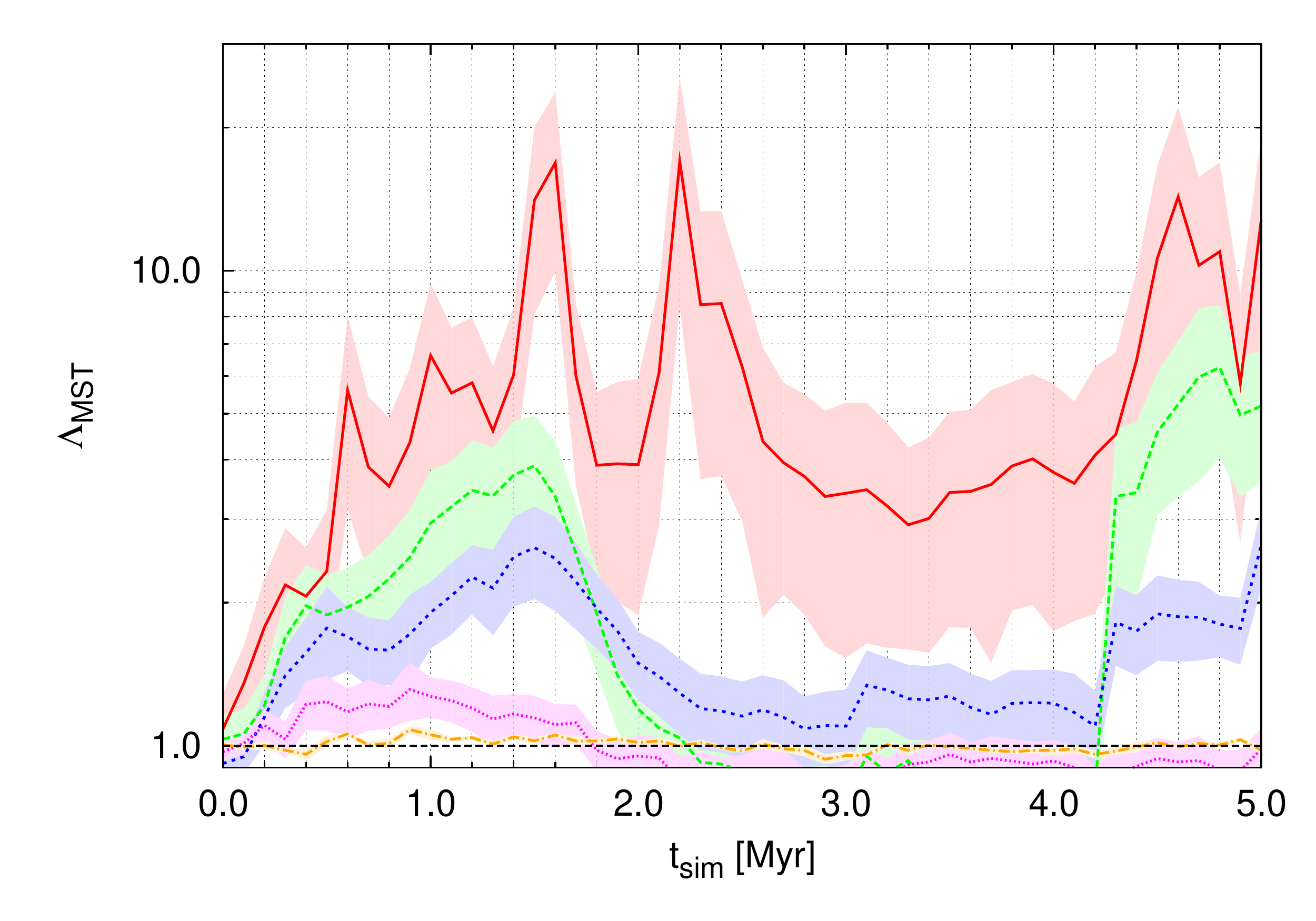}
  \caption{Dynamical mass segregation in a star cluster with 1k particles and cold initial conditions ($Q=0.1$). The plots show the mean $\Gmst$ (top)
    and $\Lmst$ (bottom) of three simulations for the 5 (red solid line), 10 (green long-dashed line), 20 (blue short-dashed line), 50 (magenta dotted
    line), and 500 (orange dot-dashed line) most massive stars over time. The filled regions indicate 1$\sigma$ uncertainties.}
  \label{fig:mass_segregation__dynamical}
\end{figure}

In Fig.~\ref{fig:mass_segregation__dynamical} we show the mean of three simulations of the same model with different random seeds used to generate
individual positions, velocities, and masses. We plot $\Gmst$ and $\Lmst$ only for the 5, 10, 20, 50, and 500 most massive stars for reasons of
visibility. The much better performance of our improved method $\Mmstg$ compared to $\Mmstl$ is evident. Though both measures do show similar maximum
degrees of mass segregation $\Lmst$ is subject to large variations over time. Only for the five most massive stars does it show high significance over
a longer period while for the 10 and 50 most massive stars even values lower than one are quite common.

In contrast, $\Gmst$ clearly increases over time with the steepest gradient within the first 2\,Myr (or 5.5 N-body time units). It clearly separates
the different mass groups, inversely correlated with the sample size. The much better robustness of $\Gmst$ reflects again its very efficient damping
of ``outliers''.

We note that mass segregation in a collapsing, intermediate-size stellar cluster of 1k stars can occur very quickly, i.e. within only a few crossing
times. This finding demonstrates that rapid mass segregation (in terms of dynamical time scale) does \emph{not} require substructure as has been
recently investigated by \citet{2009ApJ...700L..99A}. However, a detailed investigation of dynamical mass segregation in young star clusters will be
presented in an upcoming publication.

%

\section{Conclusion and Discussion}

\label{sec:conclusions}

We have developed a new method, $\Mmstg$, to measure mass segregation by significantly improving a previous approach of
\citet[][]{2009ApJ...700L..99A} based on the minimum spanning tree (MST), here referred to as $\Mmstl$. Their method uses the normalised length of the
MST of a given sample of stars, $\Lmst$, as a measure of compactness.

Compared to ``classical'' measures of mass segregation such as the slope of the (differential) mass function in different annuli around the cluster
centre (see Section~\ref{sec:introduction}), $\Lmst$ overcomes several substantial weaknesses. Our new method $\Mmstg$ inherits all the advantages
provided by $\Mmstl$ --
\begin{itemize}
\item independence of cluster geometry,
\item no requirement of cluster centre,
\item no requirement of quantitative measure of masses,
\item one-parametric unique measure of mass segregation --
\end{itemize}
and adds three significant improvements:
\begin{itemize}
\item nearly linear computational efficiency,
\item robustness against peculiar configurations, and
\item high sensitivity.
\end{itemize}

This gain is based on two boosting ingredients:
\begin{enumerate}
\item The implementation of an efficient ${\cal{O}}(N \log{N}$) algorithm for the calculation of the MST, and
\item the calculation of the geometrical mean of the connecting edges, $\Gmst$, instead of just their sum, $\Lmst$.
\end{enumerate}

The high performance is achieved by construction of a graph via 2D Delaunay triangulation \citep[from the software package
GEOMPACK:][]{1991AdvEng.13..325J}, quick sorting of its edges, and an efficient numerical implementation of Kruskal's algorithm
\citep{1956PAMS....7...48K}. The advantage of the geometrical mean is that it damps the contribution of ``outliers'' very efficiently and thus traces
the concentration of the dominant stars of a configuration.

We have demonstrated the advantage of $\Mmstg$ compared to other known and frequently used methods via various examples, both for numerical and
observational data.
\begin{enumerate}
\item In general, using only the ten to twenty most massive stars $\Gmst$ provides a robust and sensitive measure of mass segregation for the entire
  population of star clusters in our Galaxy.
\item In particular, very low degrees of mass segregation can be detected in massive clusters like NGC~3603 that consist of 10k or more stars.
\item We have also confirmed the very strong mass segregation of the five most massive stars in the ONC as reported by \citet{2006ApJ...644..355H}.
\item Incompleteness of observational data has only a moderate effect on $\Mmstg$ and corrections can be implemented easily similar to methods like
  $\Mmfd$.
\end{enumerate}

However, we also stress three important aspects that have to be considered for a careful investigation of mass segregation via $\Mmstg$:
\begin{enumerate}
\item Our conclusion that the 10 to 20 most massive stars generally provide the most sensitive measure of mass segregation has been derived from model
  clusters with smooth mass segregation over the entire sample. We caution that this is not to be expected if only specific subsamples are affected by
  mass segregation (as shown for the five most massive stars of the ONC in Section~\ref{sec:results:onc}); this is of particular concern for studies
  of primordial mass segregation. We thus generally recommend to use both the cumulative and the differential analysis of a set of mass groups to
  investigate mass segregation in (young) star clusters.
\item The signature of a small sub-sample can be strong enough to bias $\Gmst$ of a larger parent sample. To avoid wrong conclusions one should also
  calculate $\Gmst$ of disjoint samples, e.g. of the 5, 6 to 10, 11 to 20 etc. most massive stars.
\item The value of $\Gmst$ of a sub-sample is independent of its location with respect to the entire sample. It only reflects the compactness of this
  sub-sample, \emph{not} its concentration towards some centre of the entire sample. So a large $\Gmst$ does not necessarily mean that the
  corresponding sub-sample is mass-segregated.

  Thus, in practice, in particular when dealing with observational data, the determination of some centre of the sample as reference point is still
  required. However, the value of $\Gmst$ remains \emph{independent} of the reference point and so its advantage compared to other methods.
\end{enumerate}

Our improved method $\Mmstg$ will be vital for future studies of mass segregation in stellar systems. Its robustness and sensitivity is ideal for
tracing lower degrees and more complex types of mass segregation than before. This is of particular interest for the earliest stages of star formation
that seem to form protoclusters of complex structure \citep[e.g.][]{2006ApJ...636L..45T}. Furthermore, it will also help to investigate the question
whether mass segregation is observed at all in young star clusters \citep[e.g.][]{2009A&A...495..147A}. Finally, our method can be used in general for
precise structure analysis of any type of stellar system from planetary debris disks to galaxy clusters.

As a first application of $\Mmstg$ to numerical simulations we have demonstrated that mass segregation in young star clusters can occur very quickly
under dynamically cold initial conditions. Indeed we do expect young stellar systems to form under such conditions as a result of the gravitational
collapse of their parent molecular cloud. Whether collapse still occurs in largely evolved young clusters like the ONC -- the scenario corresponding
to our simulation -- is still unknown yet became favoured recently
\citep[e.g.][]{1988AJ.....95.1755J,2000NewA....4..615K,2005MNRAS.358..742S,2008ApJ...676.1109F,2009ApJ...697.1020P,2009ApJ...697.1103T}. However, at
least some fraction of the oldest stars must have formed during global collapse. Hence rapid mass segregation affects the structure of star forming
regions at the earliest stages and could thus explain the observed mass segregation of the most massive stars in the ONC.

We will investigate this and other aspects of dynamical mass segregation in young star clusters under various conditions in an upcoming publication.

%

\begin{acknowledgements}
  CO and RS acknowledge support by NAOC CAS through the Silk Road Project, and by Global Networks and Mobility Program of the University of Heidelberg
  (ZUK 49/1 TP14.8 Spurzem). CO appreciates funding by the German Research Foundation (DFG), grant OL~350/1-1. RS is funded by the Chinese Academy of
  Sciences Visiting Professorship for Senior International Scientists, Grant Number 2009S1-5. We have partly used the special supercomputers at the
  Center of Information and Computing at National Astronomical Observatories, Chinese Academy of Sciences, funded by Ministry of Finance of People’s
  Republic of China under the grant ZDY Z2008−2.

  We thank S. Aarseth for providing the highly sophisticated N-body code $\nbody$ (and its GPU extension) and greatly appreciate his support.

\end{acknowledgements}

%

\bibliographystyle{aa}
\bibliography{references}

\begin{thebibliography}{39}
\expandafter\ifx\csname natexlab\endcsname\relax\def\natexlab#1{#1}\fi

\bibitem[{{Aarseth}(2003)}]{2003grav.book.....A}
{Aarseth}, S. 2003, {Gravitational N-body Simulations} (Cambridge, Cambridge
  University Press, 2003, 430 p.)

\bibitem[{{Allison} {et~al.}(2009{\natexlab{a}}){Allison}, {Goodwin}, {Parker},
  {de Grijs}, {Portegies Zwart}, \& {Kouwenhoven}}]{2009ApJ...700L..99A}
{Allison}, R.~J., {Goodwin}, S.~P., {Parker}, R.~J., {et~al.}
  2009{\natexlab{a}}, \apjl, 700, L99

\bibitem[{{Allison} {et~al.}(2009{\natexlab{b}}){Allison}, {Goodwin}, {Parker},
  {Portegies Zwart}, {de Grijs}, \& {Kouwenhoven}}]{2009MNRAS.395.1449A}
{Allison}, R.~J., {Goodwin}, S.~P., {Parker}, R.~J., {et~al.}
  2009{\natexlab{b}}, \mnras, 395, 1449

\bibitem[{{Ascenso} {et~al.}(2009){Ascenso}, {Alves}, \&
  {Lago}}]{2009A&A...495..147A}
{Ascenso}, J., {Alves}, J., \& {Lago}, M.~T.~V.~T. 2009, \aap, 495, 147

\bibitem[{{Bolte}(1989)}]{1989ApJ...341..168B}
{Bolte}, M. 1989, \apj, 341, 168

\bibitem[{{Cartwright} \& {Whitworth}(2004)}]{2004MNRAS.348..589C}
{Cartwright}, A. \& {Whitworth}, A.~P. 2004, \mnras, 348, 589

\bibitem[{{Evans} {et~al.}(2009){Evans}, {Dunham}, {J{\o}rgensen}, {Enoch},
  {Mer{\'{\i}}n}, {van Dishoeck}, {Alcal{\'a}}, {Myers}, {Stapelfeldt},
  {Huard}, {Allen}, {Harvey}, {van Kempen}, {Blake}, {Koerner}, {Mundy},
  {Padgett}, \& {Sargent}}]{2009ApJS..181..321E}
{Evans}, N.~J., {Dunham}, M.~M., {J{\o}rgensen}, J.~K., {et~al.} 2009, \apjs,
  181, 321

\bibitem[{{Farouki} {et~al.}(1983){Farouki}, {Hoffman}, \&
  {Salpeter}}]{1983ApJ...271...11F}
{Farouki}, R.~T., {Hoffman}, G.~L., \& {Salpeter}, E.~E. 1983, \apj, 271, 11

\bibitem[{{F{\H u}r{\'e}sz} {et~al.}(2008){F{\H u}r{\'e}sz}, {Hartmann},
  {Megeath}, {Szentgyorgyi}, \& {Hamden}}]{2008ApJ...676.1109F}
{F{\H u}r{\'e}sz}, G., {Hartmann}, L.~W., {Megeath}, S.~T., {Szentgyorgyi},
  A.~H., \& {Hamden}, E.~T. 2008, \apj, 676, 1109

\bibitem[{Gower \& Ross(1969)}]{1969JRSS...18...54G}
Gower, J.~C. \& Ross, G. J.~S. 1969, Journal of the Royal Statistical Society.
  Series C (Applied Statistics), 18, pp. 54

\bibitem[{{Hillenbrand}(1997)}]{1997AJ....113.1733H}
{Hillenbrand}, L.~A. 1997, \aj, 113, 1733

\bibitem[{{Hillenbrand} \& {Hartmann}(1998)}]{1998ApJ...492..540H}
{Hillenbrand}, L.~A. \& {Hartmann}, L.~W. 1998, \apj, 492, 540

\bibitem[{{Huff} \& {Stahler}(2006)}]{2006ApJ...644..355H}
{Huff}, E.~M. \& {Stahler}, S.~W. 2006, \apj, 644, 355

\bibitem[{{Jeffries}(2007)}]{2007MNRAS.376.1109J}
{Jeffries}, R.~D. 2007, \mnras, 376, 1109

\bibitem[{{Joe}(1991)}]{1991AdvEng.13..325J}
{Joe}, B. 1991, Advances in Engineering Software, 13, 325

\bibitem[{{Jones} \& {Walker}(1988)}]{1988AJ.....95.1755J}
{Jones}, B.~F. \& {Walker}, M.~F. 1988, \aj, 95, 1755

\bibitem[{{Khalisi} {et~al.}(2007){Khalisi}, {Amaro-Seoane}, \&
  {Spurzem}}]{2007MNRAS.374..703K}
{Khalisi}, E., {Amaro-Seoane}, P., \& {Spurzem}, R. 2007, \mnras, 374, 703

\bibitem[{{Kraus} {et~al.}(2009){Kraus}, {Weigelt}, {Balega}, {Docobo},
  {Hofmann}, {Preibisch}, {Schertl}, {Tamazian}, {Driebe}, {Ohnaka}, {Petrov},
  {Sch{\"o}ller}, \& {Smith}}]{2009A&A...497..195K}
{Kraus}, S., {Weigelt}, G., {Balega}, Y.~Y., {et~al.} 2009, \aap, 497, 195

\bibitem[{{Kroupa}(2000)}]{2000NewA....4..615K}
{Kroupa}, P. 2000, New Astronomy, 4, 615

\bibitem[{{Kroupa}(2001)}]{2001MNRAS.322..231K}
{Kroupa}, P. 2001, \mnras, 322, 231

\bibitem[{Kruskal(1956)}]{1956PAMS....7...48K}
Kruskal, Joseph~B., J. 1956, Proceedings of the American Mathematical Society,
  7, pp. 48

\bibitem[{{Lada} \& {Lada}(2003)}]{2003ARA&A..41...57L}
{Lada}, C.~J. \& {Lada}, E.~A. 2003, \araa, 41, 57

\bibitem[{{McCaughrean} \& {Stauffer}(1994)}]{1994AJ....108.1382M}
{McCaughrean}, M.~J. \& {Stauffer}, J.~R. 1994, \aj, 108, 1382

\bibitem[{{Menten} {et~al.}(2007){Menten}, {Reid}, {Forbrich}, \&
  {Brunthaler}}]{2007A&A...474..515M}
{Menten}, K.~M., {Reid}, M.~J., {Forbrich}, J., \& {Brunthaler}, A. 2007, \aap,
  474, 515

\bibitem[{{Olczak} {et~al.}(2010){Olczak}, {Pfalzner}, \&
  {Eckart}}]{2010A&A...509A..63O}
{Olczak}, C., {Pfalzner}, S., \& {Eckart}, A. 2010, \aap, 509, A260000+

\bibitem[{{Pandey} {et~al.}(1992){Pandey}, {Mahra}, \&
  {Sagar}}]{1992BASI...20..287P}
{Pandey}, A.~K., {Mahra}, H.~S., \& {Sagar}, R. 1992, Bulletin of the
  Astronomical Society of India, 20, 287

\bibitem[{{Preparata} \& {Shamos}(1985)}]{1985cgai.book.....P}
{Preparata}, F.~P. \& {Shamos}, M.~I. 1985, {Computational geometry. an
  introduction} (Springer New York Inc., 1985)

\bibitem[{{Proszkow} {et~al.}(2009){Proszkow}, {Adams}, {Hartmann}, \&
  {Tobin}}]{2009ApJ...697.1020P}
{Proszkow}, E.-M., {Adams}, F.~C., {Hartmann}, L.~W., \& {Tobin}, J.~J. 2009,
  \apj, 697, 1020

\bibitem[{{Richer} {et~al.}(1988){Richer}, {Fahlman}, \&
  {Vandenberg}}]{1988ApJ...329..187R}
{Richer}, H.~B., {Fahlman}, G.~G., \& {Vandenberg}, D.~A. 1988, \apj, 329, 187

\bibitem[{{Scally} {et~al.}(2005){Scally}, {Clarke}, \&
  {McCaughrean}}]{2005MNRAS.358..742S}
{Scally}, A., {Clarke}, C., \& {McCaughrean}, M.~J. 2005, \mnras, 358, 742

\bibitem[{{Schmeja} \& {Klessen}(2006)}]{2006A&A...449..151S}
{Schmeja}, S. \& {Klessen}, R.~S. 2006, \aap, 449, 151

\bibitem[{{Spitzer}(1969)}]{1969ApJ...158L.139S}
{Spitzer}, L.~J. 1969, \apjl, 158, L139+

\bibitem[{{Spurzem} \& {Takahashi}(1995)}]{1995MNRAS.272..772S}
{Spurzem}, R. \& {Takahashi}, K. 1995, \mnras, 272, 772

\bibitem[{{Stolte} {et~al.}(2006){Stolte}, {Brandner}, {Brandl}, \&
  {Zinnecker}}]{2006AJ....132..253S}
{Stolte}, A., {Brandner}, W., {Brandl}, B., \& {Zinnecker}, H. 2006, \aj, 132,
  253

\bibitem[{Tarjan(1979)}]{1979JACM...26..690T}
Tarjan, R.~E. 1979, J. ACM, 26, 690

\bibitem[{{Teixeira} {et~al.}(2006){Teixeira}, {Lada}, {Young}, {Marengo},
  {Muench}, {Muzerolle}, {Siegler}, {Rieke}, {Hartmann}, {Megeath}, \&
  {Fazio}}]{2006ApJ...636L..45T}
{Teixeira}, P.~S., {Lada}, C.~J., {Young}, E.~T., {et~al.} 2006, \apjl, 636,
  L45

\bibitem[{{Tobin} {et~al.}(2009){Tobin}, {Hartmann}, {Furesz}, {Mateo}, \&
  {Megeath}}]{2009ApJ...697.1103T}
{Tobin}, J.~J., {Hartmann}, L., {Furesz}, G., {Mateo}, M., \& {Megeath}, S.~T.
  2009, \apj, 697, 1103

\bibitem[{{{\v S}ubr} {et~al.}(2008){{\v S}ubr}, {Kroupa}, \&
  {Baumgardt}}]{2008MNRAS.385.1673S}
{{\v S}ubr}, L., {Kroupa}, P., \& {Baumgardt}, H. 2008, \mnras, 385, 1673

\bibitem[{{Yu} {et~al.}(2011){Yu}, {de Grijs}, \& {Chen}}]{2011arXiv1103.0406Y}
{Yu}, J., {de Grijs}, R., \& {Chen}, L. 2011, ArXiv e-prints

\end{thebibliography}

%

\appendix

\section{Data incompleteness and $\Gmst$}

\label{app:incompleteness}

We provide an example on how incompleteness of observational data affects $\Gmst$ and present a simple prescription for recovering the original
signature of mass segregation via the corresponding completeness function.

Here we use again the original sample of 929 stars in the ONC from \citet{1997AJ....113.1733H} to generate a reduced sample via convolution with an
artificial completeness function with radial and mass dependence of the form
\begin{multline}
  \label{eq:completeness}
  c( r, r_{\mathrm{u}}, p, a_{\mathrm{l}}, a_{\mathrm{u}}, m, m_{\mathrm{l}}, m_{\mathrm{u}}, b_{\mathrm{l}}, b_{\mathrm{u}} ) = \\
  1 - \exp \left\{ {\cal{R}}( r, r_{\mathrm{u}}, p, a_{\mathrm{l}}, a_{\mathrm{u}} ) \left[ {\cal{M}}(m, m_{\mathrm{l}}, m_{\mathrm{u}})
    \right]^{{\cal{R}}(r, r_{\mathrm{u}}, p, b_{\mathrm{l}}, b_{\mathrm{u}} ) } \right\} \,,
\end{multline}
where 
\begin{align}
  {\cal{R}}( r, r_{\mathrm{u}}, p, q_{\mathrm{l}}, q_{\mathrm{u}} ) &= ( q_{\mathrm{u}} - q_{\mathrm{l}} ) ( r / r_{\mathrm{u}} )^p +
  q_{\mathrm{l}} \,,\\
  {\cal{M}}( m, m_{\mathrm{l}}, m_{\mathrm{u}} ) &= \frac{ \log( m / m_{\mathrm{l}} ) } { \log( m / m_{\mathrm{u}} ) } \,,
\end{align}
and the indices $l$ and $u$ denote lower and upper values, respectively.

\begin{figure}
  \centering
  \includegraphics[height=0.95\linewidth,angle=-90]{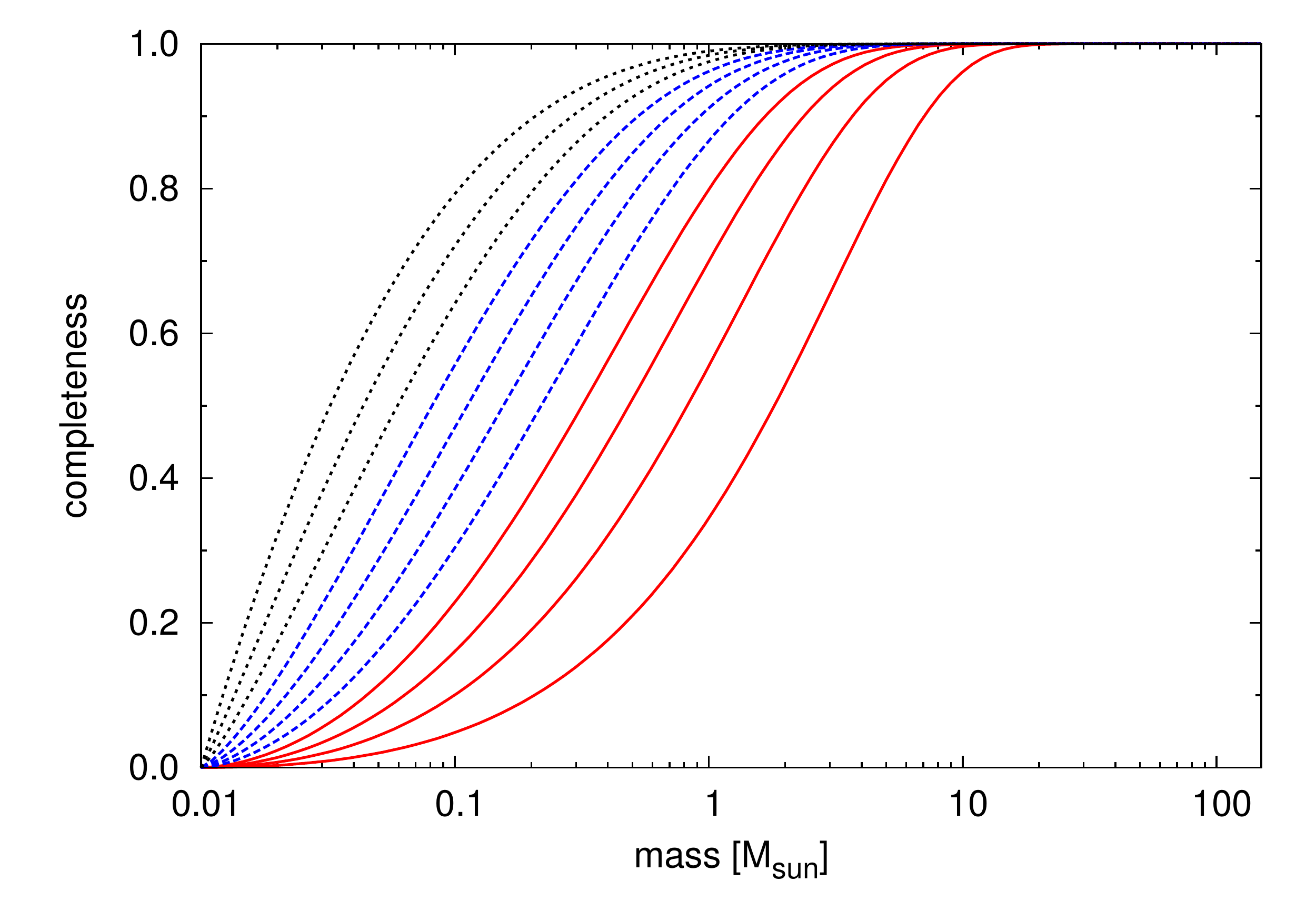}
  \caption{Model of completeness as a function of stellar mass and radial position in the ONC according to Eq.~\eqref{eq:completeness} and using the
    parameters from Table~\ref{tab:completeness_parameters}. From right to left the graphs represent increasing radii in steps of 0.25\,pc from the
    centre to 2.5\,pc, the outer boundary of the ONC. The solid (red), dashed (blue), and dotted (black) lines are grouped in steps of 1\,pc,
    respectively.}
  \label{fig:completeness}
\end{figure}

The parameters used to simulate loss of data due to observational incompleteness are presented in Table~\ref{tab:completeness_parameters}. The
corresponding plot in Fig.~\ref{fig:completeness} shows the completeness as a function of stellar mass in dependence of radial position in the
cluster. From right to left the graphs represent increasing radii in steps of 0.25\,pc from the centre to 2.5\,pc, the outer boundary of the
ONC. The solid (red), dashed (blue), and dotted (black) lines are grouped in steps of 1\,pc, respectively.

The functional form has been adopted such that it resembles observational incompleteness due to crowding in the dense and bright cluster centre and
due to individual extinction over the entire cluster. It is expected that stars with masses $\gtrsim 10\,\Msun$ are too bright to be missed while
low-mass objects close to the cluster centre with it's four luminous Trapezium stars remain mostly undetected.

\begin{table}
  \centering
  \begin{tabular}{*{8}{c}}
    $r_{\mathrm{u}}$ & $p$ & $a_{\mathrm{l}}$ & $a_{\mathrm{u}}$ & $m_{\mathrm{l}}$ & $m_{\mathrm{u}}$ & $b_{\mathrm{l}}$ & $b_{\mathrm{u}}$ \\
    \hline
    \hline
    2.5              & 1   & 0.5              & 5                & 0.01             & 150              & 1                & 2
  \end{tabular}
  \caption{Set of parameters used for Eq.~\eqref{eq:completeness} in the present study.}
  \label{tab:completeness_parameters}
\end{table}

To reconstruct a larger unbiased sample from the incomplete sample we have used the inverse completeness function in the following simple
prescription:
\begin{enumerate}
\item Go to the next object $i$ in the incomplete sample.
\item Calculate its completeness $c_i$ using its radial position $r_i$ and mass $m_i$.
\item If $c_i > c_{\mathrm{min}}$:
  \begin{enumerate}
  \item Calculate the missing number of objects $n_i = 1/c_i - 1$ with $r_i$ and $m_i$.
  \item Generate for each additional object $j$ its
    \begin{enumerate}
    \item mass $m_j = m_i ( 1 + \lambda )$, $\lambda \in [-0.05, 0.05]$ a random number, and
    \item radial position $r_j = r_i ( 1 + \eta )$, $\eta \in [-0.1, 0.1]$ a random number.
    \end{enumerate}
  \end{enumerate}
\end{enumerate}

In short, the parents' individual completeness $c_i$ determines the number of ``clones'' $n_i$ with slightly varying properties. The threshold
$c_{\mathrm{min}}$ is used to prevent massive cloning of rare objects that could introduce a significant bias. We found that a surprisingly low value
of $c_{\mathrm{min}} = 0.2$ is save.

\begin{figure}
  \centering
  \includegraphics[height=0.95\linewidth,angle=-90]{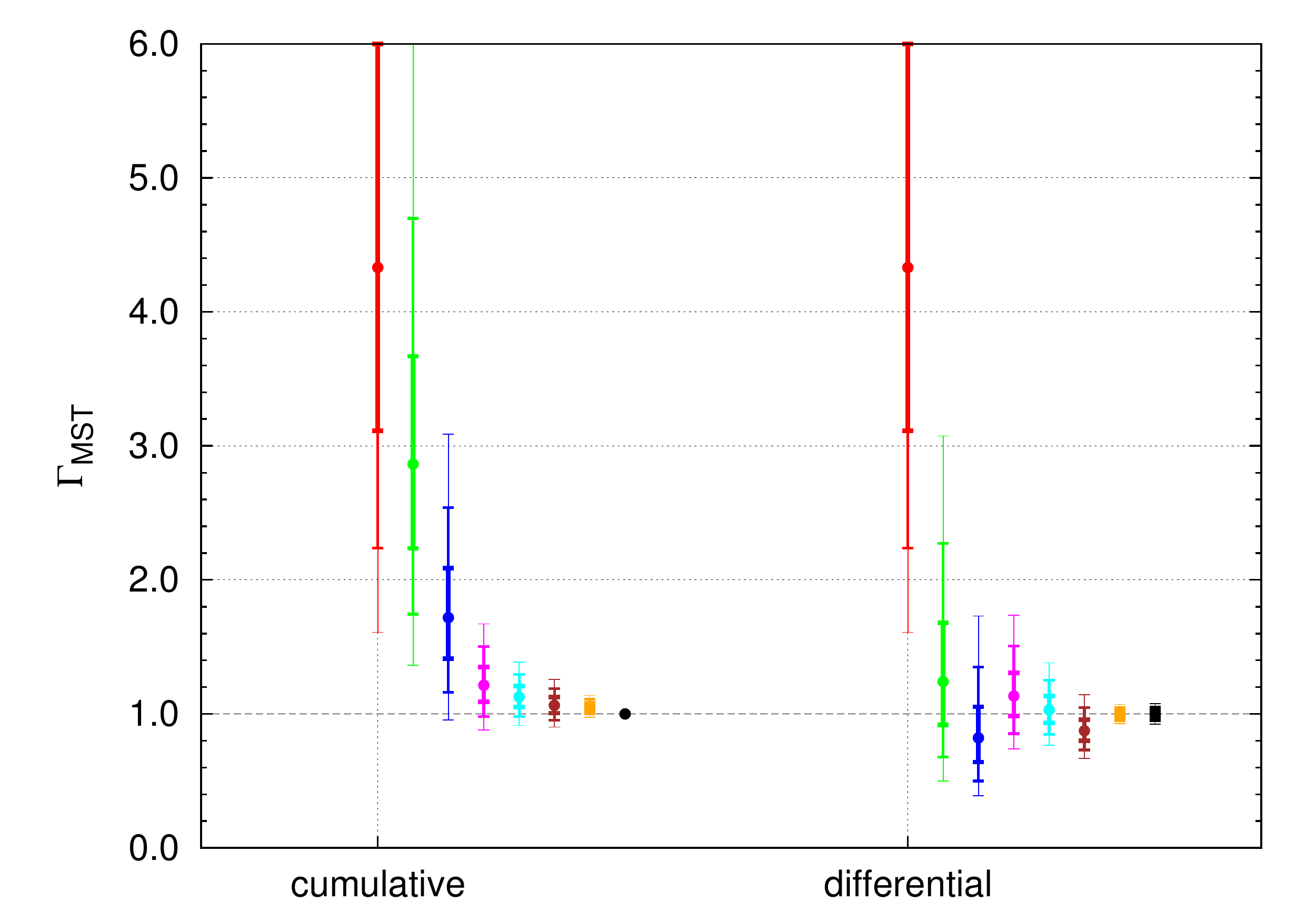}
  \includegraphics[height=0.95\linewidth,angle=-90]{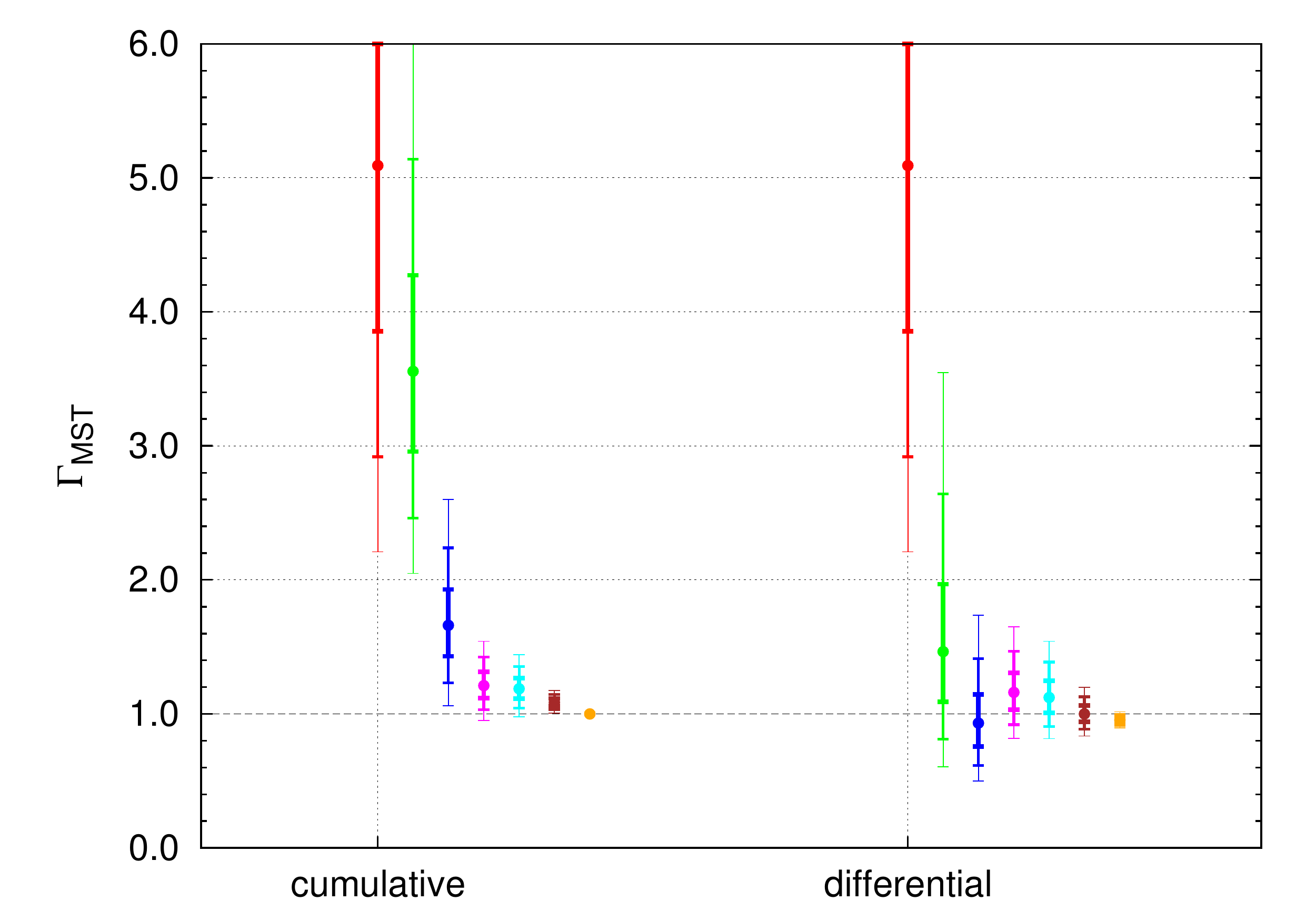}
  \includegraphics[height=0.95\linewidth,angle=-90]{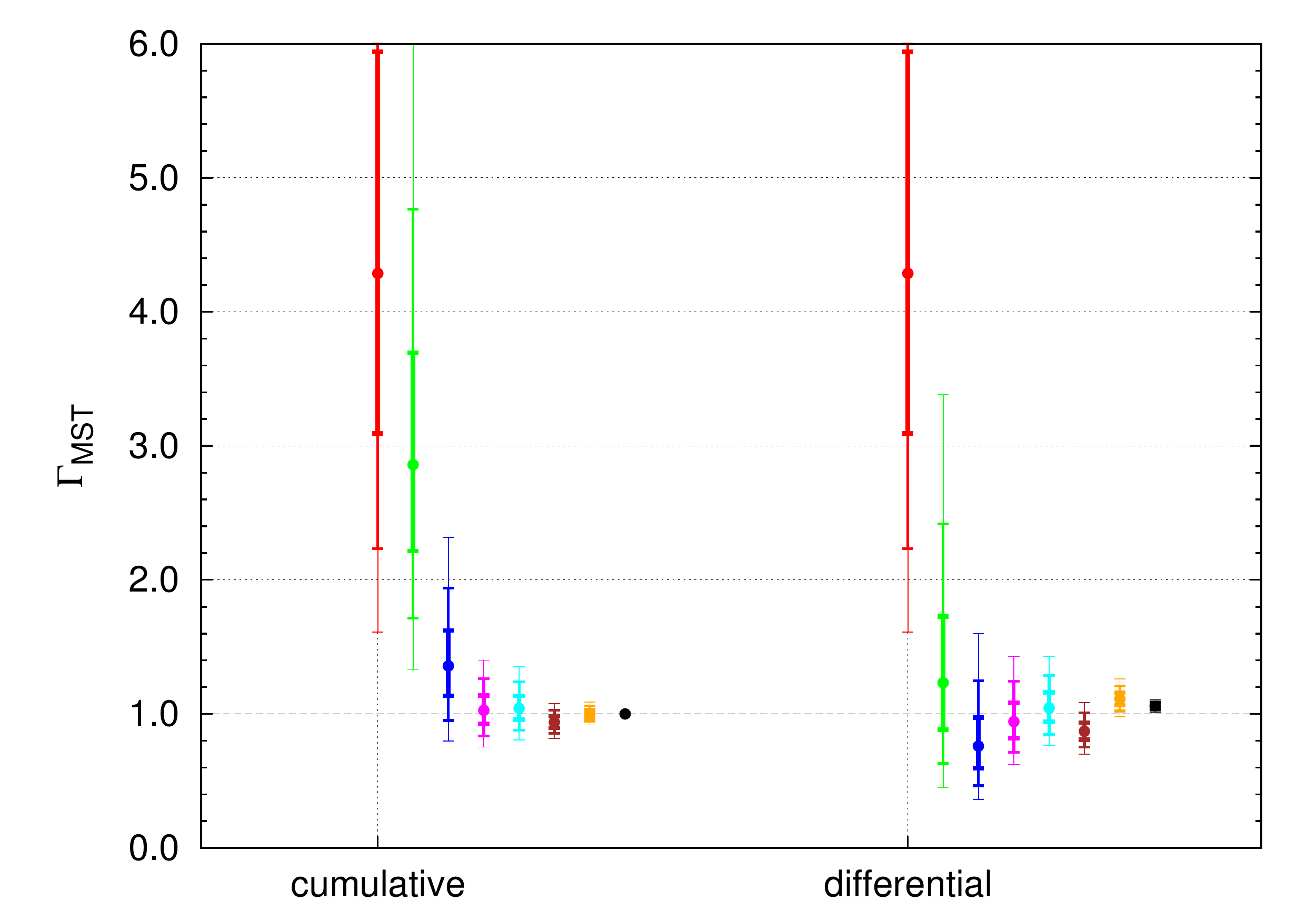}
  \caption{Diagnostics of mass segregation for three sets of observational data of the young Orion Nebula Cluster (ONC).  The left-hand side shows
    from left to right $\Gmst$ of the 5, 10, 20, 50, 100, 200, 500 most massive, and all stars, i.e. cumulative mass groups; the right-hand side shows
    from left to right $\Gmst$ of the 5, 6 to 10, 11 to 20, 21 to 50, 51 to 100, 101 to 200, 201 to 500, and 500 to all most massive stars,
    i.e. differential mass groups. The error bars and line thickness mark the $1\sigma$, $2\sigma$, and $3\sigma$ uncertainties. \emph{Top:} Original
    set of 929 stars observed by \citet{1997AJ....113.1733H}. \emph{Middle:} Incomplete sample. \emph{Bottom:} Reconstructed sample.}
  \label{fig:mass_segregation__onc__incompleteness}
\end{figure}

We compare $\Gmst$ of the three samples -- original, incomplete, and reconstructed -- in Fig.~\ref{fig:mass_segregation__onc__incompleteness}.  The
left-hand side shows $\Gmst$ of the 5, 10, 20, 50, 100, 200, 500 most massive, and all stars, i.e. cumulative mass groups; the right-hand side shows
$\Gmst$ of the 5, 6 to 10, 11 to 20, 21 to 50, 51 to 100, 101 to 200, 201 to 500, and 500 to all most massive stars, i.e. differential mass
groups. The top plot refers to the original sample of 929 ONC stars: it is the same data as on the left-hand side of
Fig.~\ref{fig:mass_segregation__onc}. Below we plot $\Gmst$ of the incomplete sample that has been artificially reduced to 485 stars via
Eq.~\eqref{eq:completeness}. This removal of nearly half the cluster population -- preferentially low-mass stars in the cluster centre -- has a
marginal effect on $\Gmst$: only for the five and ten most massive stars a significant difference is observed. However, the differential plot reveals
again a dominant contribution from the five most massive stars, $\Gmst$ of the 6-10 most massive remains nearly unchanged.

The bottom plot contains data of additional 345 cloned objects and so 830 stars in total -- about 10\,\% less than the original sample. The
reconstruction procedure seems to work fairly well: $\Gmst$ of the five and ten most massive stars is indistinguishable from its original
values. However, the next intervals up to the 200 most massive stars all show slightly lower degrees of mass segregation than originally. The reason
is that the sample of even less massive stars (200-500 and 500-830 most massive stars) is marginally mass segregated (i.e. $\Gmst > 1$ unlike in the
original sample): these low-mass stars that make up the main part of the random reference sample were cloned preferentially closer to the cluster
centre which reduced $\Gmst$ of the more massive stars.

However, considering the reconstruction of up to 80\,\% individual incompleteness and nearly half of the entire incomplete sample, the result is very
promising: the strong conclusion for the original sample that the five most massive stars in the ONC are highly mass segregated while all other stars
do not show any signature of mass segregation remains unchanged for the reconstructed sample.

%

\end{document}